\documentclass[a4paper,12pt]{article}
\usepackage{authblk}
\usepackage{array}
\usepackage{booktabs}
\usepackage{graphicx}
\usepackage{cite}
\usepackage{authblk}
\title{ Determination of $\alpha$-optical potential for reactions with p-nuclei from the study of ($\alpha$,n) reactions in the astrophysically relevant energy region }
\author[1,2]{Dipali Basak}
\author[1]{Chinmay Basu}
\affil[1]{Nuclear Physics Division, Saha Institute of Nuclear Physics,1/AF, Bidhannagar, Kolkata-700064}
\affil[2]{Homi Bhabha National Institute, Anushaktinagar, Mumbai, Maharashtra 400094}
\date{}                     
\newcommand{\lambdabar}{{\mkern0.75mu\mathchar '26\mkern -9.75mu\lambda}}
\begin{document}
\maketitle

\abstract{
Optical potential parameters in nuclear model calculations are determined by fitting elastic scattering angular distribution data. Due to the dominance of Coulomb part, elastic scattering is performed at much higher energies. A different approach using ($\alpha$,n) reaction is suggested to determine the alpha optical potentials suitable for astrophysically relevant low energy regions. Reaction on p-nuclei in the  mass region A $\approx$ 92-168 have been chosen and a modified McFadden-Satchler $\alpha$-optical potential is obtained from fitting the ($\alpha$,n) reaction data. The effect of level density and $\gamma$-ray strength function is also studied using this new potential by analyzing ($\alpha,\gamma$) cross-sections.
\section{Introduction}
\label{intro}
Most of the heavy nuclei are synthesized in stars by neutron capture competing with $\beta$-decays in s- or r-process. In contrast to the nuclei formed from s- or r-process, 30-35 stable neutron deficient nuclei from $^{74}$Se-$^{196}$Hg (so-called p-nuclei) are synthesized by the $\gamma$-process\cite{a,b,c}. The $\gamma$-process is different combinations of the ($\gamma$,n),(($\gamma$,p) and ($\gamma,\alpha$) reactions at high temperature($\sim$ GK). For $\gamma$-process calculations, a huge reaction network involving about 1000 stable or unstable nuclei is required. Reaction rates obtained from reaction cross-sections using Hauser-Feshbach (HF) statistical model calculations around the Gamow window region are important ingredients for network calculation. These reaction cross-sections of the $\gamma$-process are calculated from inverse capture reactions using the principle of detailed balance. HF calculations are sensitive to the nuclear physics input parameters (Optical potentials, Level densities,$\gamma$-ray strength function, etc) and a systematic study with respect to reactions involving p-nuclei is required.\\
 Several capture reactions on p-nuclei have been performed in recent years. HF calculations for proton and neutron capture reactions satisfactorily explain the measured cross-section\cite{d,e,f,g,h,i,j,k,l,m,n,o,p,q}. But significant differences were observed in the measured cross-section data compared to the HF predictions of alpha capture cross-section\cite{r,s,t,u,v,w,x,y}.These differences mainly come from the choice of inappropriate entrance channel alpha optical model potential(AOMP).\\
In this work, ($\alpha$,n) reaction cross-sections were compared with the HF calculation and a modified AOMP was proposed for reactions with p-nuclei.
 It is also observed that besides AOMP, other input parameters like level density (LD), $\gamma$-ray strength function ($\gamma$SF) play an important role in statistical model calculations and affect the theoretical cross-section data of the $(\alpha,\gamma)$ reactions.

\section{Theoretical formalism}
Integrated alpha induced compound nucleus cross-sections for a specific channel \textit{x} ( \textit{x} = n,$\alpha$,p,$\gamma$, etc ) is given by the Hauser-Feshbach theory as
\begin{equation}
\sigma(\alpha,x)= \sum_{J} \sigma(E_{c},J) P(E_{c},J;x)
\end{equation} 
Where $\sigma(E_{c},J)$ is the compound nucleus formation cross-section. 
The total integrated (over emission energy and angle) decay probability  to a particular channel \textit{x} is given by 
\begin{equation}
P(E_{c},J;x) = \frac{R(E_c,J;x)dE}{R(E_c,J)dE}
\end{equation}
The decay rates for particle $\left(R_{p}(E_c,J)\right)$ and $\gamma$-channel $\left(R_{\gamma}(E_c,J\right)$ are given 
\begin{equation}
R_{p}(E_c,J)dE = \sum_{x} \sum_{j,s}\int_{\varepsilon=0}^{E_c-S_x}R_{x}\left(E_c,J;E_c-S_x-\varepsilon,j,s\right)d\varepsilon
\end{equation}
\begin{equation}
R_{\gamma}(E_c,J)dE = \sum_{\lambda} \sum_{j}\int_{\varepsilon=0}^{E_c}R_{\lambda}\left(E_c,J;E_c-\varepsilon,j\right)d\varepsilon
\end{equation}
 The total decay rate including both particles and $\gamma$-rays is given by
 \begin{equation}
 R\left(E_c,J\right) = R_{\gamma}(E_c,J) + R_{p}(E_c,J)
 \end{equation}
 The particle and $\gamma$-decay rates from the initial state ($E_c$, J) to the final state ($U, j$) are given by Eq.\ref{Eq6} and Eq.\ref{Eq7} and are dependent on the level density ($\rho$) and $\gamma$-ray strength functions($c_{\lambda}$ ).
\begin{equation}
R_{\gamma}(E_c,J;U,j)dE = \left[  c_{\lambda}(\varepsilon)\right] \left[\varepsilon^{2\lambda+1}\right] \left[\frac{\rho(U,j)}{\rho(E_c,J)}\right]dE \label{Eq6}
\end{equation} 
Where $\varepsilon = E_c$ - U and $\vec{J} = \vec{\lambda} + \vec{j}$, $\lambda$ = multi-polarity of $\gamma$-ray.
\begin{equation}
R_{x}(E_c,J;U,j)dE = \frac{1}{h} \sum_{S=\mid j-s \mid}^{j+s}\sum_{l=J-S}^{J+S}T_l(\varepsilon) \left[\frac{\rho(U,j)}{\rho(E_c,J)}\right]dE \label{Eq7}
\end{equation} 
  Where $\varepsilon = E_c$ - U - S$_x$  , S$_x$ = Separation energy for particle \textit{x} with orbital angular momentum \textit{l} and spin \textit{s}. The transmission coefficients $T_l(\varepsilon)$ are determined from nuclear optical potentials.\\
 The expression for level density is given by 
 \begin{equation}
 \rho\left(U,j\right) = \frac{1}{2}\frac{2j+1}{2\sqrt{2\pi}\sigma^3}exp\left[-\frac{(j+\frac{1}{2})^2}{2\sigma^2}\right]\frac{\sqrt{\pi}}{12}\frac{exp(2\sqrt{aU})}{a^{\frac{1}{4}}U^{\frac{5}{4}}}
 \end{equation}
Where $\sigma^2$ is the spin cut-off parameter and \textit{a} is the level density parameter. \\
Above neutron threshold energy, neutron decay dominates over charge particle decay ($\alpha$, p, etc.) due to the absence of the Coulomb barrier. So, decay probability($P(E_{c},J;x)$) becomes unity for neutron emission and ($\alpha$,n) reaction cross-section only depends on compound nucleus formation cross-section $\sigma(E_{c},J)$. 
\begin{equation}
 \sigma(\alpha,n)\sim\sum_{J}\sigma(E_{c},J) =\sum_{J} \pi\lambdabar_{\alpha}^2T_{l}(\alpha)
 \label{Eq9}
 \end{equation}
 
$T_{l}(\alpha)$ is the alpha transmission coefficient and is calculated using Wood-Saxon optical potential is defined as
\begin{equation}
 U(r) = V_C(r) + \frac{-V_0}{1 + exp\left(\frac{r-r_vA_T^{1/3}}{a_v}\right)} +i\frac{-W_0}{1 + exp\left(\frac{r-r_wA_T^{1/3}}{a_w}\right)}
 \end{equation}
 $V_C$(r) is the coulomb potential. $V_0$ and $W_0$ are the potential depth of the real and imaginary parts of nuclear potential respectively. $r_i$ and $a_i$ (i=v,w) are the radii and diffusivities.\\
 Since the entrance channel involves only the $\alpha$-particle transmission coefficient (Eq.\ref{Eq9}), the ($\alpha$,n) cross-section is solely dependent on the $\alpha$-optical potential that determines $T_{l}(\alpha)$.
However, sensitivity studies\cite{z} show that some reactions on p-nuclei behave differently from this approximation. This is exhibited in the sensitivity plots given in Fig.\ref{FIG:1}. In the measured energy range, the ($\alpha$,n) reaction cross-sections for these nuclei are not only sensitive to $\alpha$-width but also sensitive to n and $\gamma$-width. Therefore, it is not appropriate to search for $\alpha$-optical potential from ($\alpha$,n) reaction data for these nuclei. The ($\alpha$,n) reactions on p-nuclei that are mainly sensitive to $\alpha$-width are that involve $^{92,94}$Mo, $^{96}$Ru, $^{108}$Cd, and $^{130}$Ba nuclei. Contributions from the width of other than that $\alpha$-width are $<$25$\%$. The motivation of this work lies in studying how far these non $\alpha$-widths affect the calculations and the process of extracting the $\alpha$-optical potential from ($\alpha$,n) reactions.

\begin{figure*}
\begin{center}
\begin{tabular}{cc}
\includegraphics[scale=0.24]{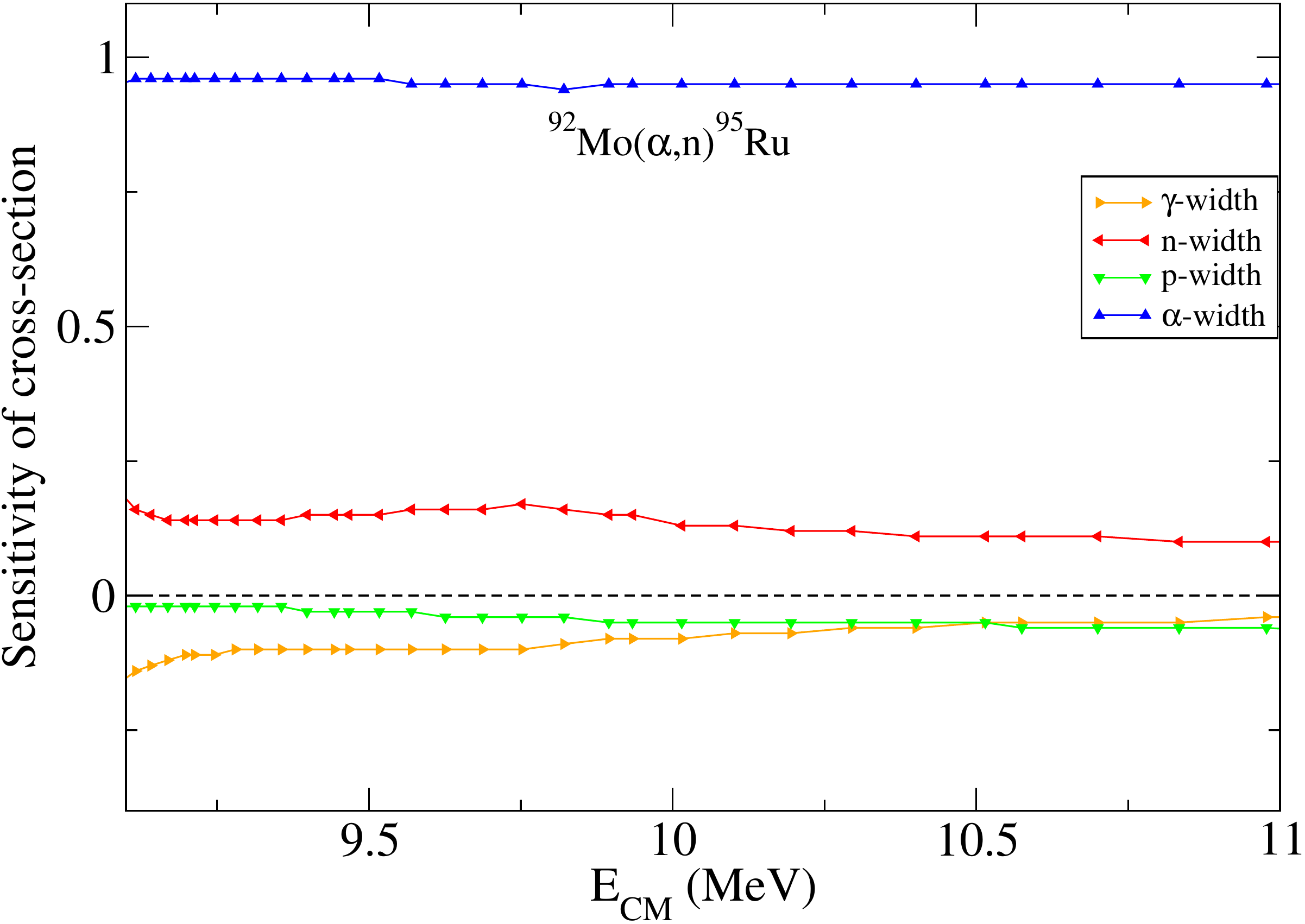}
\includegraphics[scale=0.24]{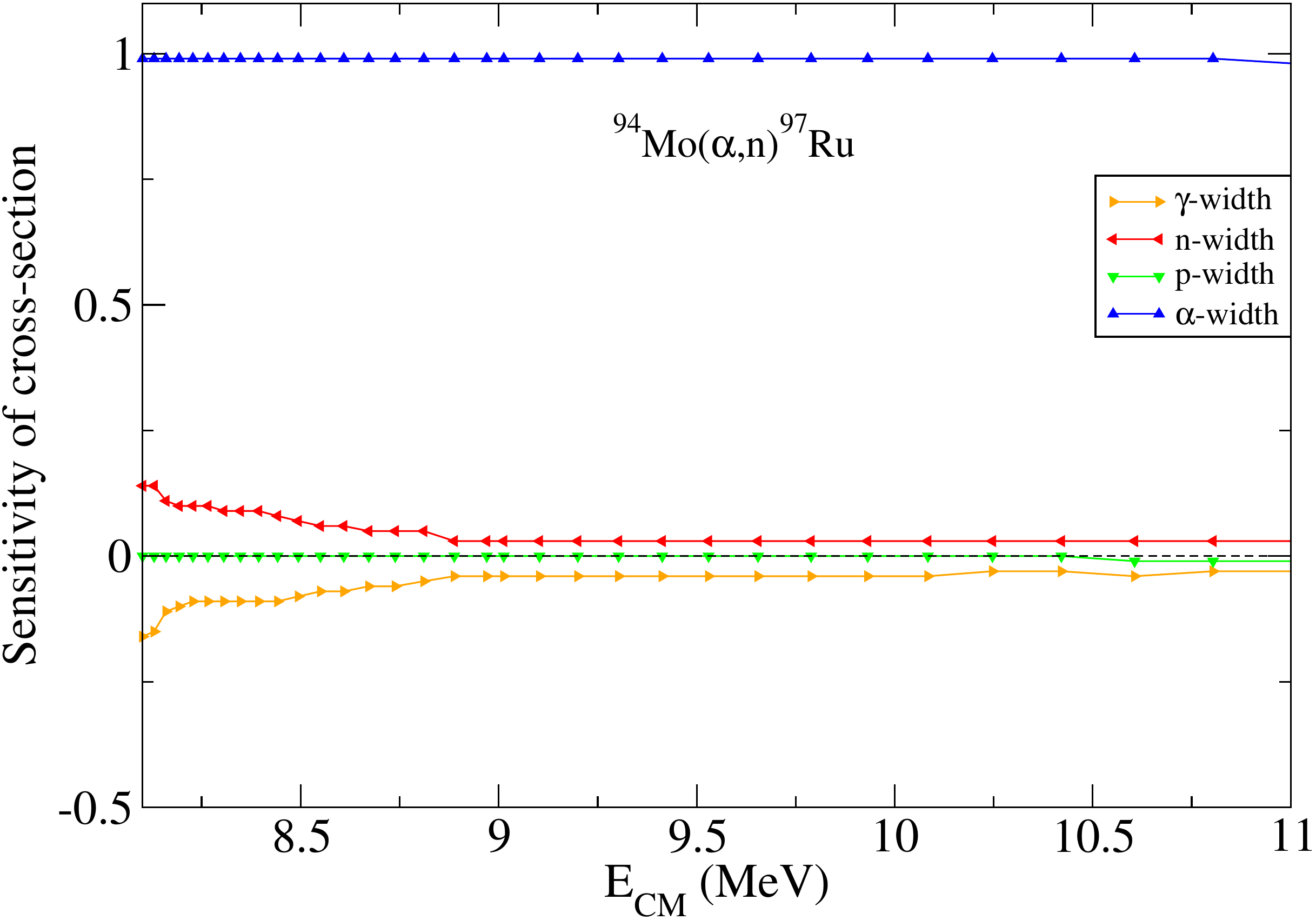}\\
\includegraphics[scale=0.24]{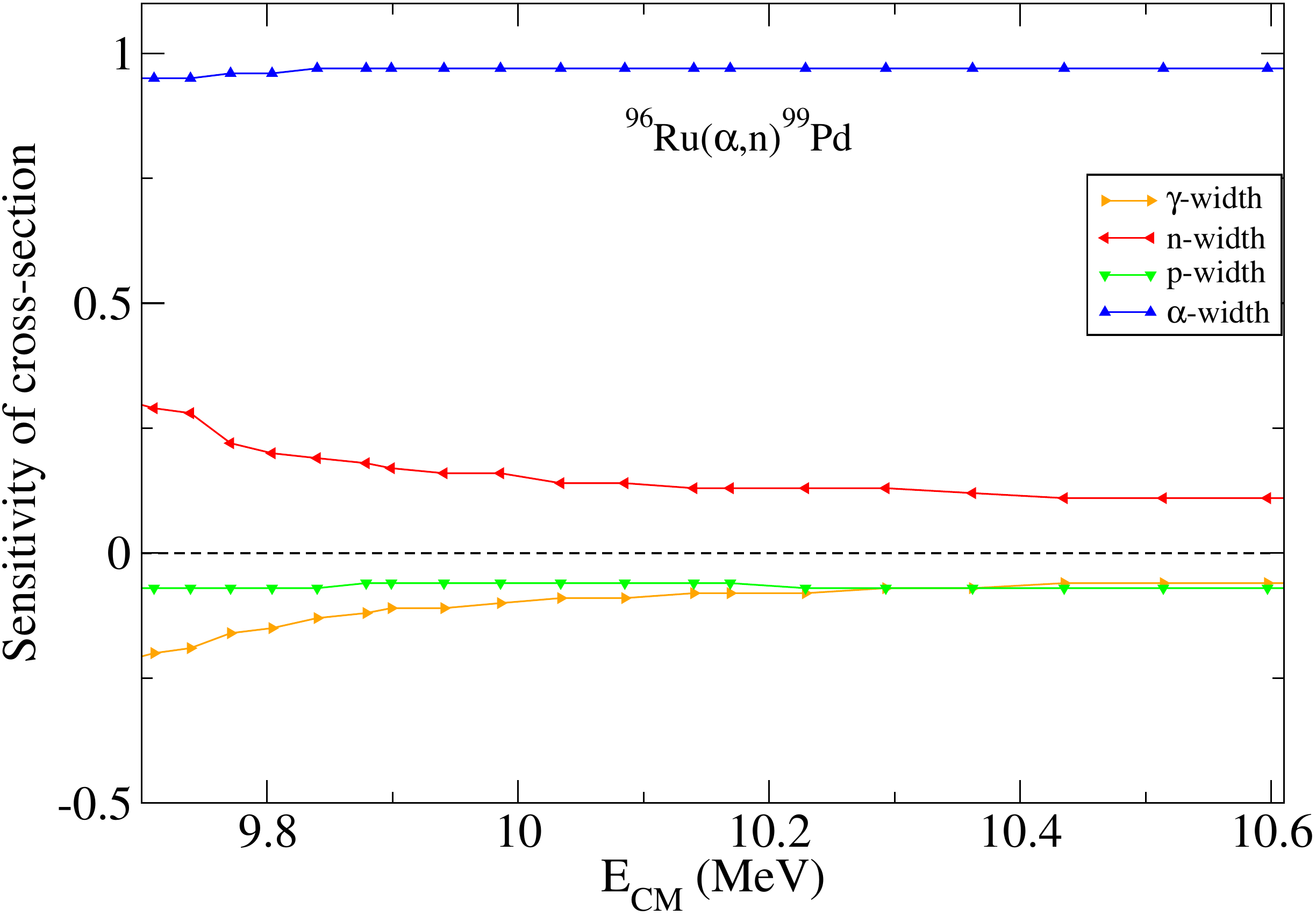}
\includegraphics[scale=0.24]{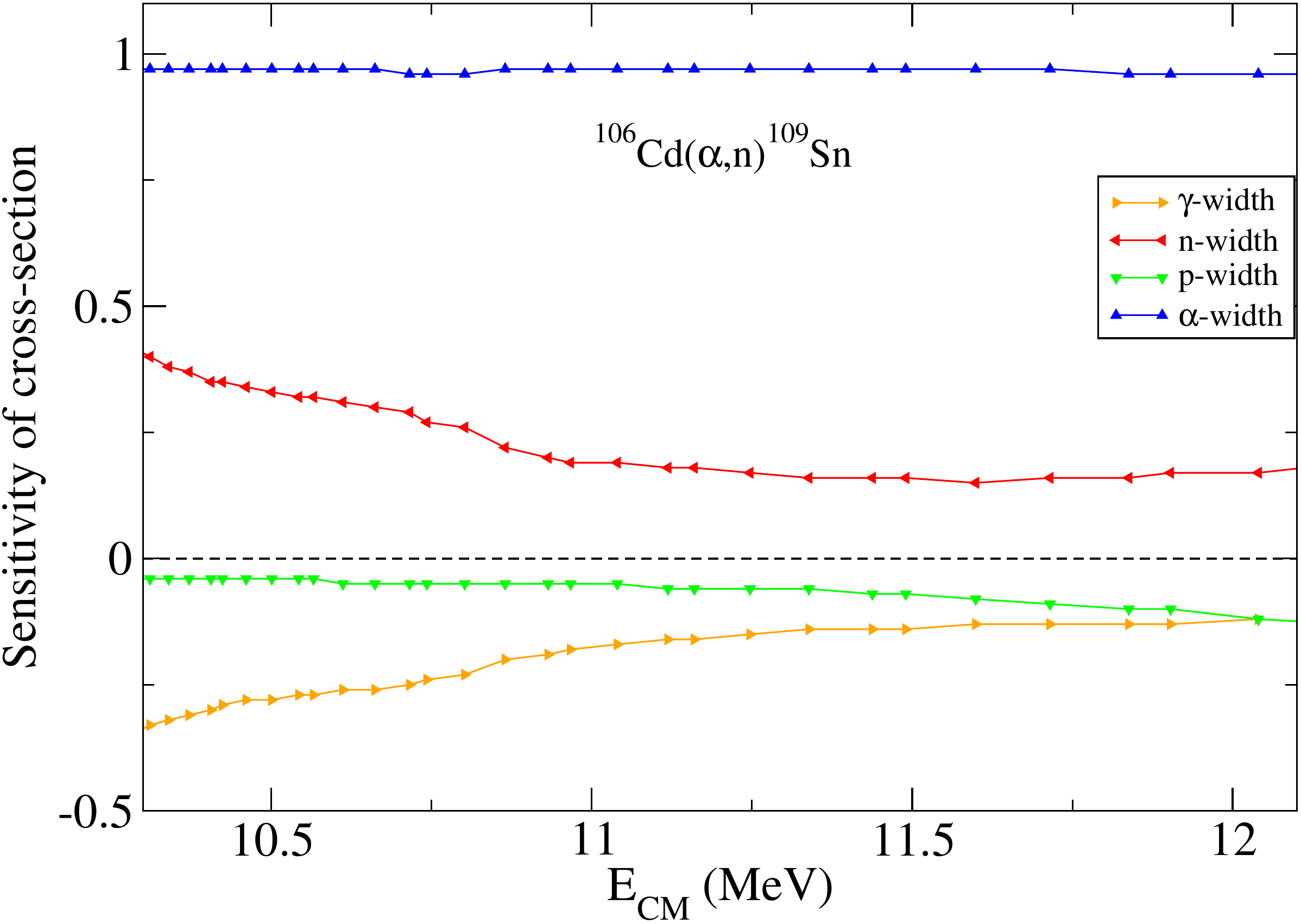}\\
\includegraphics[scale=0.24]{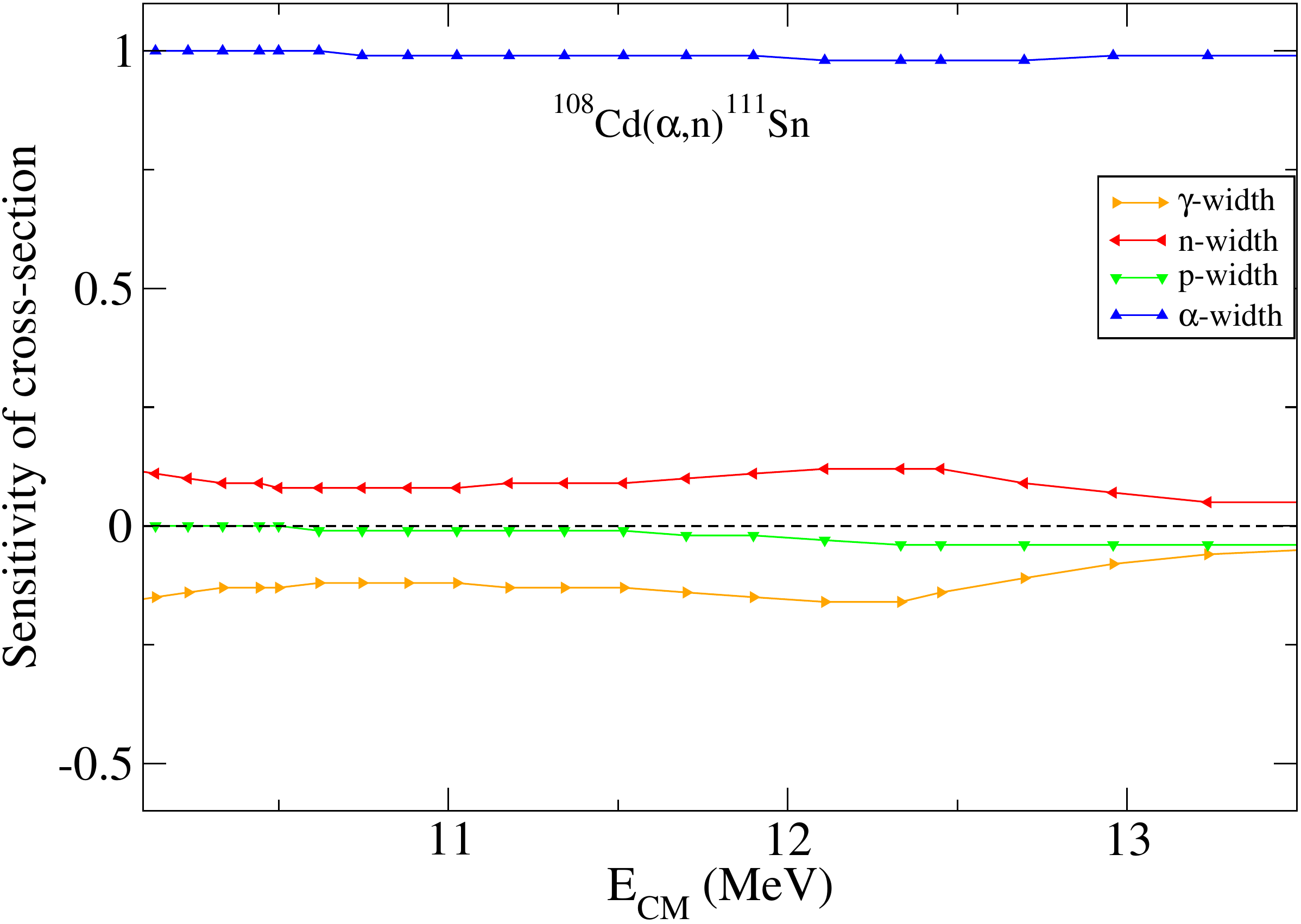}
\includegraphics[scale=0.24]{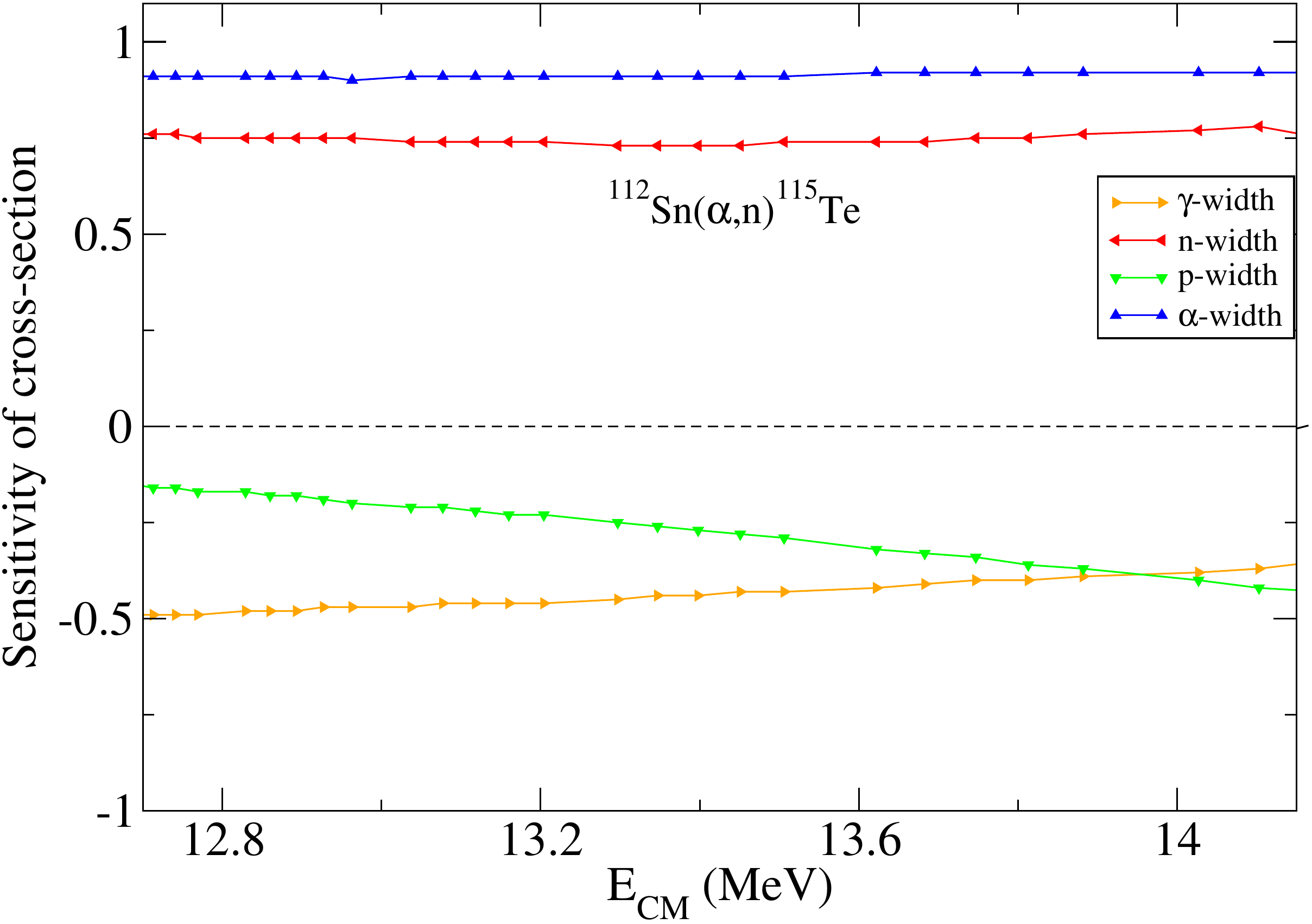}\\
\includegraphics[scale=0.24]{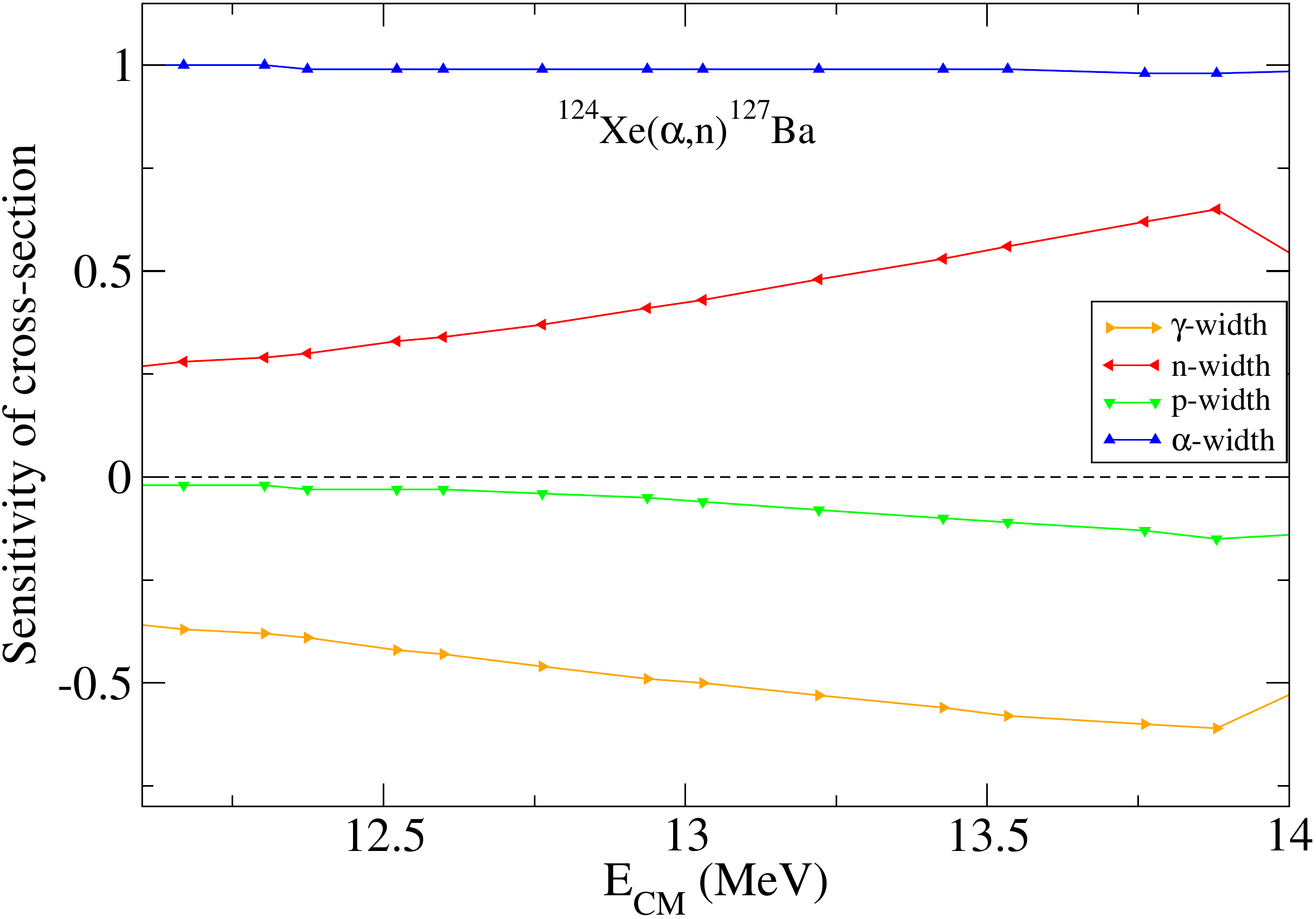}
\includegraphics[scale=0.24]{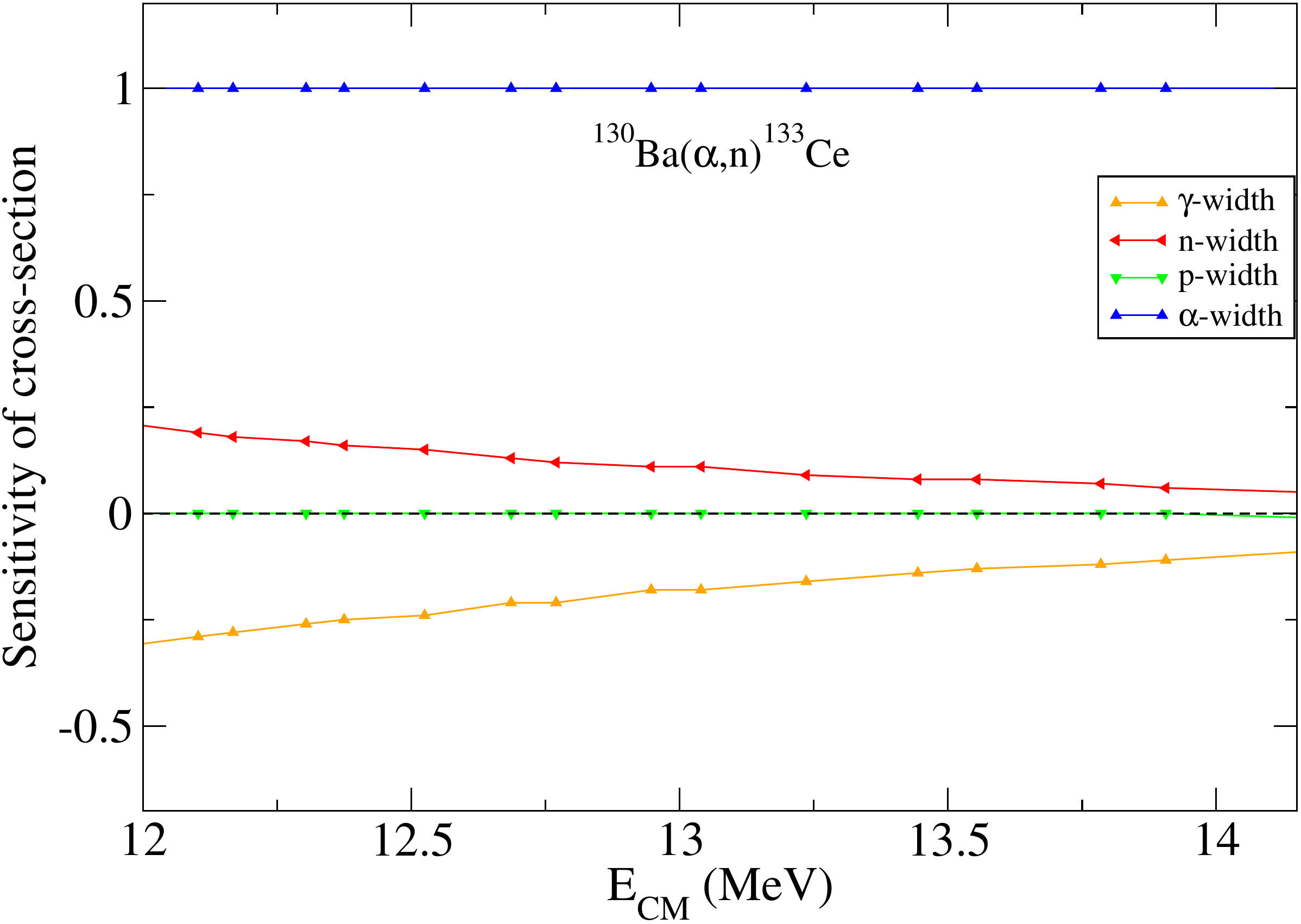}\\
\includegraphics[scale=0.24]{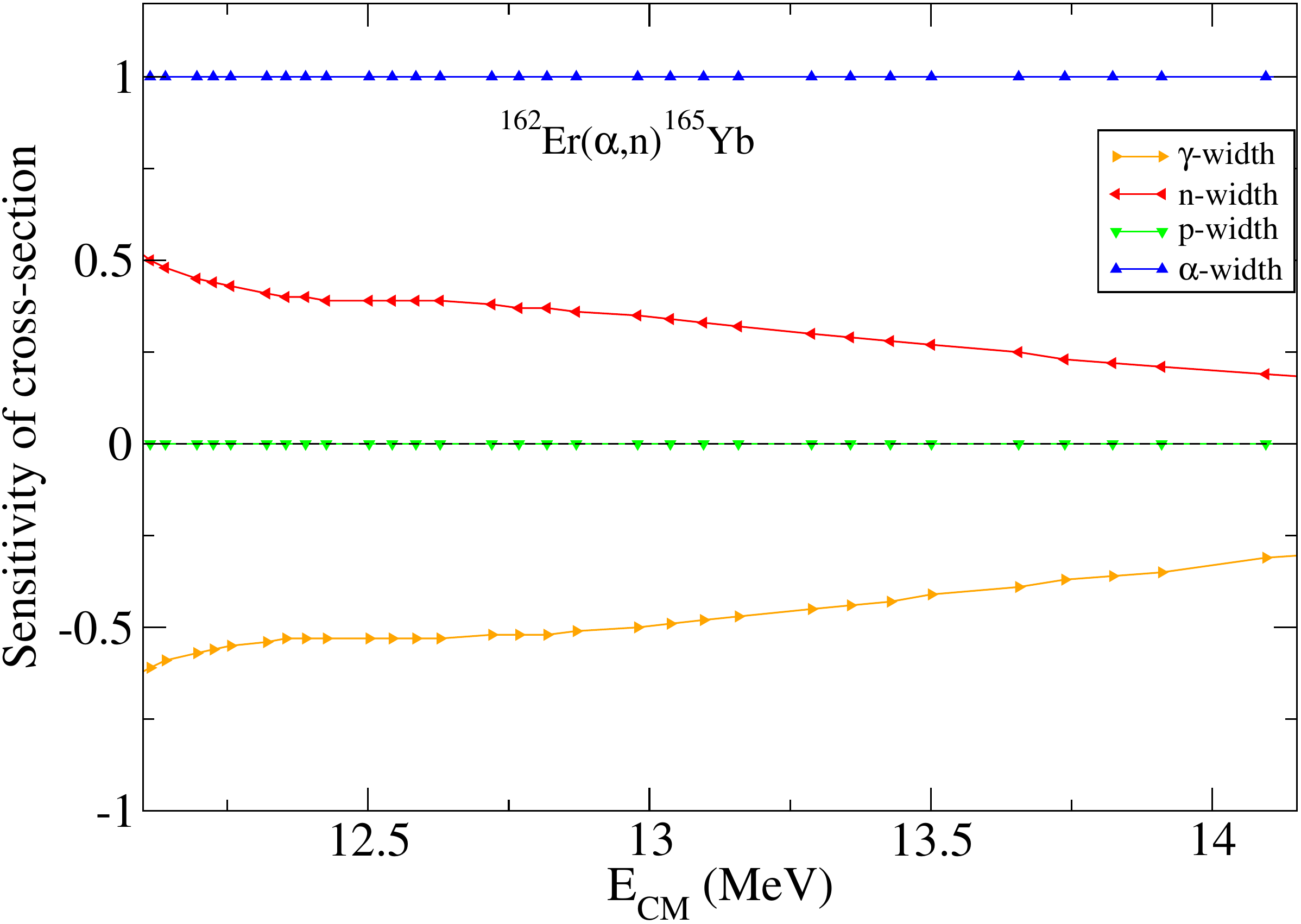}
\includegraphics[scale=0.24]{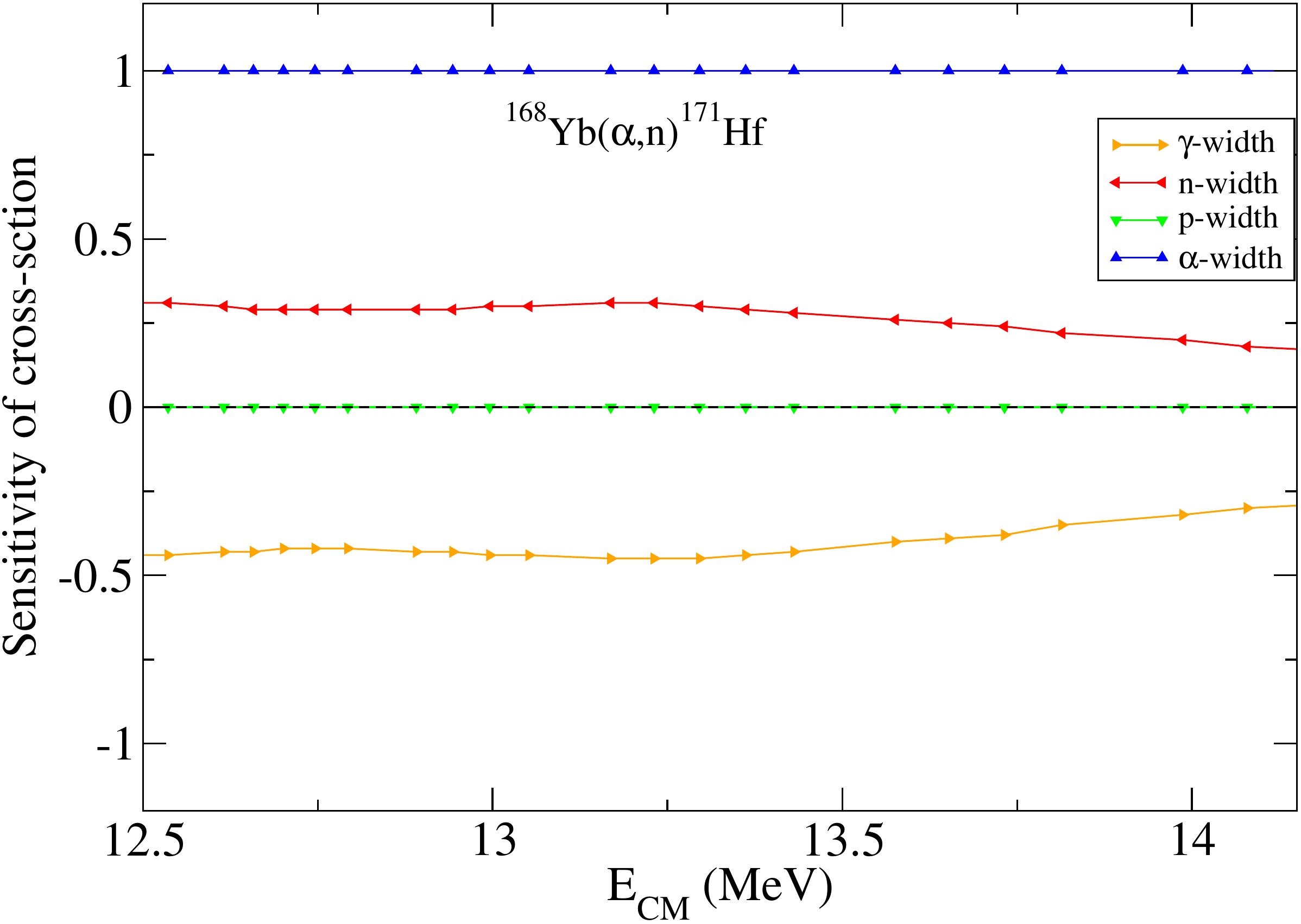}\\
\end{tabular}
\end{center}
\caption{Sensitivity of ($\alpha$,n) reaction cross-section for $^{92,94}$Mo, $^{96}$Ru, $^{106,108}$Cd, $^{112}$Sn, $^{124}$Xe, $^{130}$Ba, $^{162}$Er, and $^{168}$Yb with variation of $\alpha$-,n-,p- and $\gamma$-width\cite{z}}
\label{FIG:1}       
\end{figure*}

\section{Results and Calculations}
 \subsection{($\alpha$,n) reaction}
The HF calculations are carried out using the statistical model code TALYS in version 1.95 \cite{a1}. The dependence of the calculations on the optical model potential parameters, level density parameters, and E1 $\gamma$-ray strength functions are studied. The calculations are compared with experimental cross-sections for ($\alpha$,n) and ($\alpha,\gamma$) reactions on ten different even-even p-nuclei viz. $^{92,94}$Mo, $^{96}$Ru, $^{106,108}$Cd, $^{112}$Sn, $^{124}$Xe, $^{130}$Ba, $^{162}$Er, $^{168}$Yb.\\
 As default, AOMP in the TALYS is the Avrigeanu potential\cite{b1} for entrance channel, constant temperature Fermi gas model(CTFGM)\cite{c1} for level density, and Brink-Axel Lorentzian(BAL)\cite{d1,e1} for E1 $\gamma$-ray strength function. The results for the ($\alpha$,n) reaction are shown in Fig.\ref{FIG:2}. 
At first, the results using global AOMP  viz. that of McFadden-Satchler (McF) \cite{f1}, Avrigeanu \cite{b1} and Demetriou \cite{g1} are used in the framework of TALYS and results are found to be satisfactory for $^{94}$Mo, $^{106}$Cd, $^{130}$Ba nuclei but overestimated for the other nuclei (Fig.\ref{FIG:2}). The McF was chosen for modification to obtain satisfactory results for the cases of $^{92}$Mo, $^{96}$Ru, $^{108}$Cd, $^{112}$Sn, and $^{124}$Xe, as McF is simple energy and mass independent potential. McF potential was modified with energy dependent Fermi function on the depth parameters (both real and imaginary).

\begin{figure*}
\begin{center}
\begin{tabular}{cc}
\includegraphics[scale=0.24]{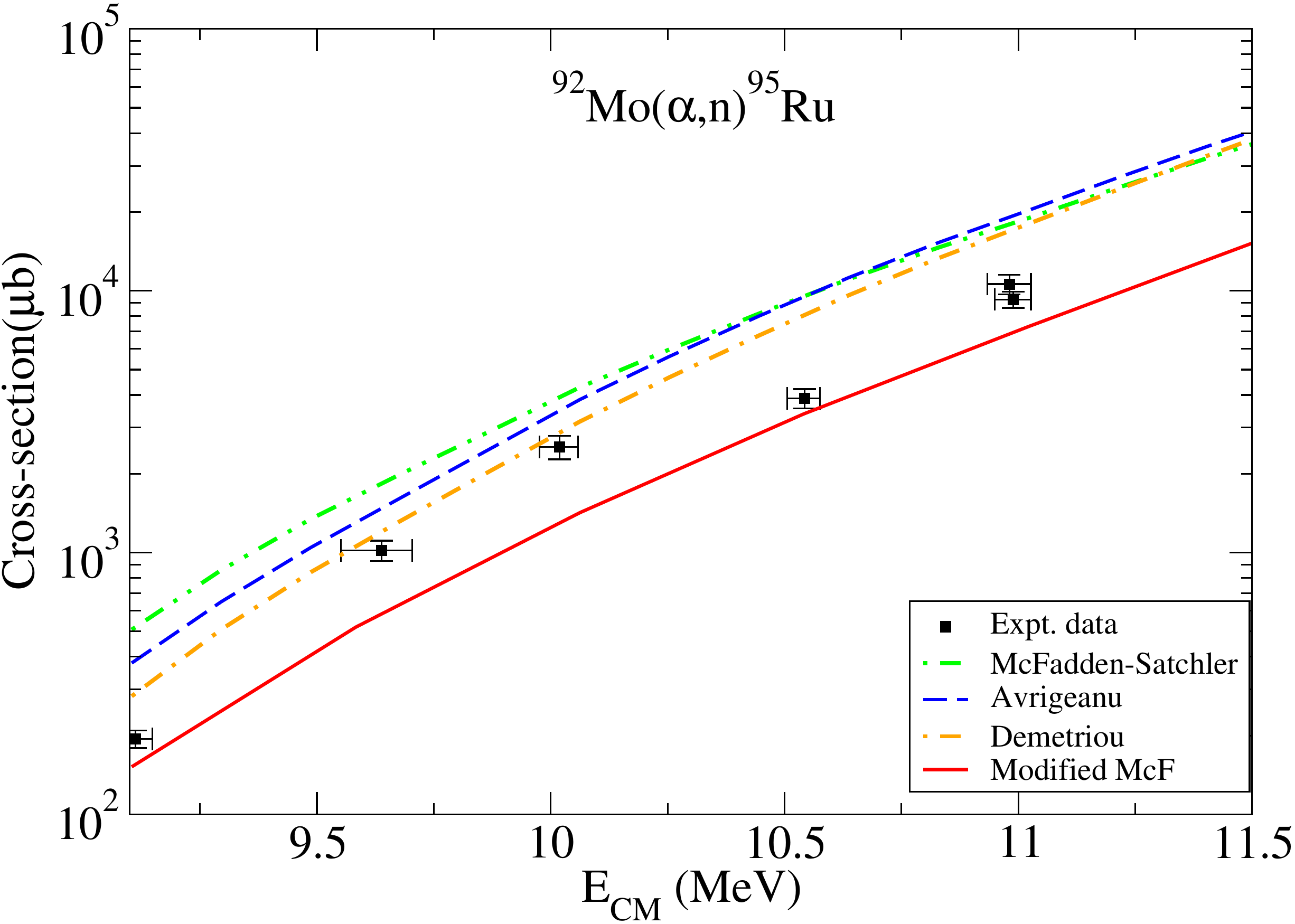}
\includegraphics[scale=0.24]{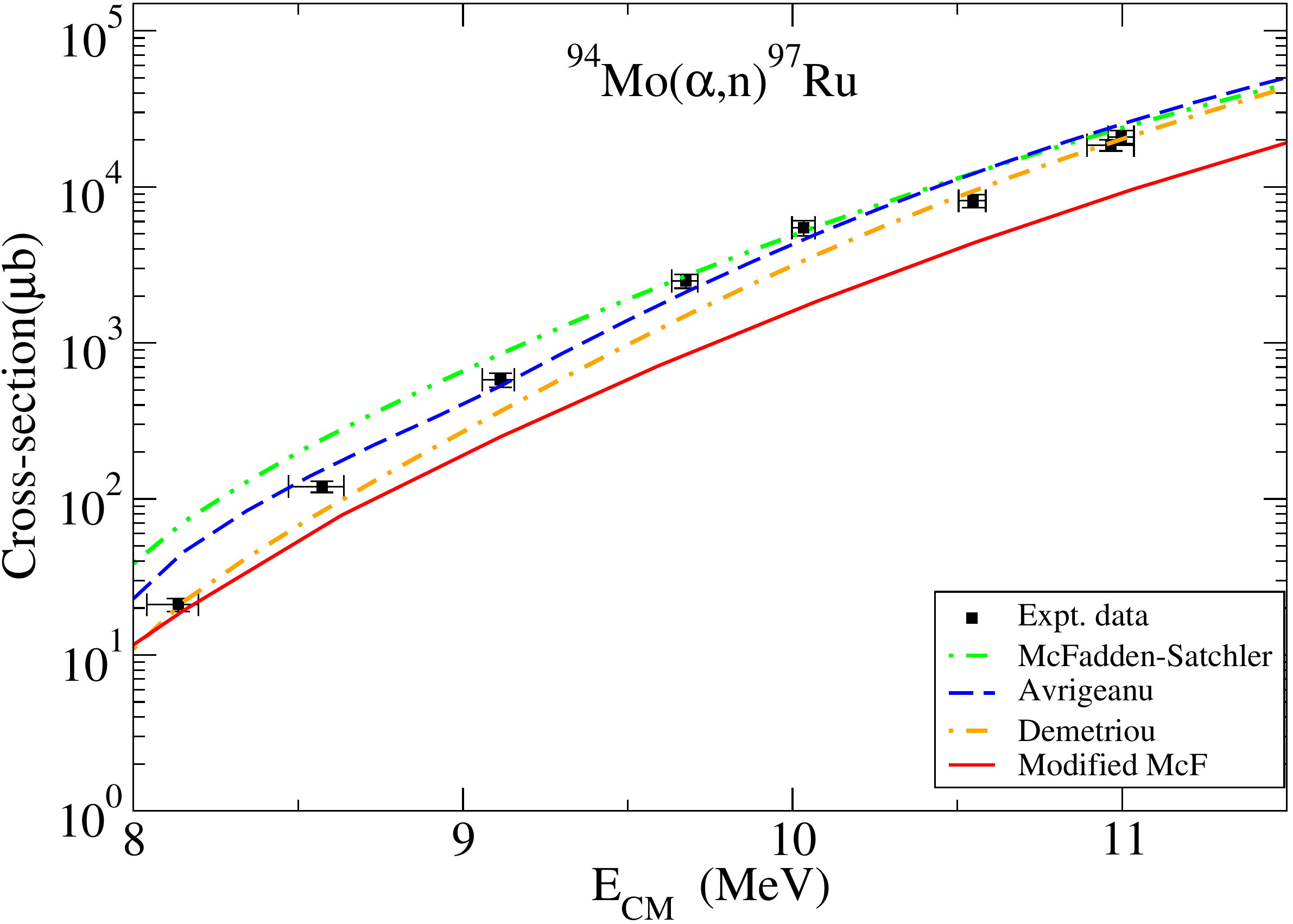}\\
\includegraphics[scale=0.24]{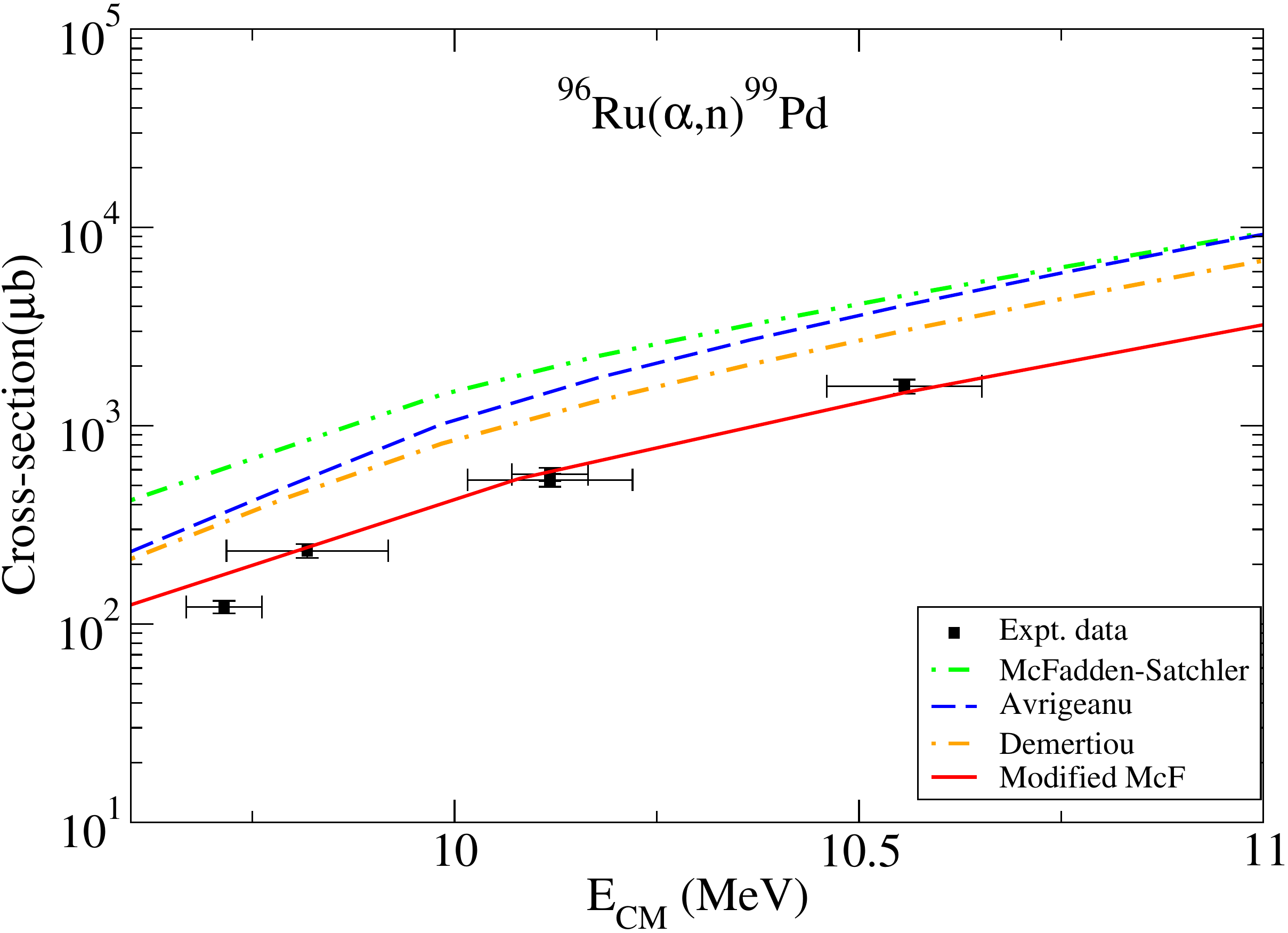}
\includegraphics[scale=0.24]{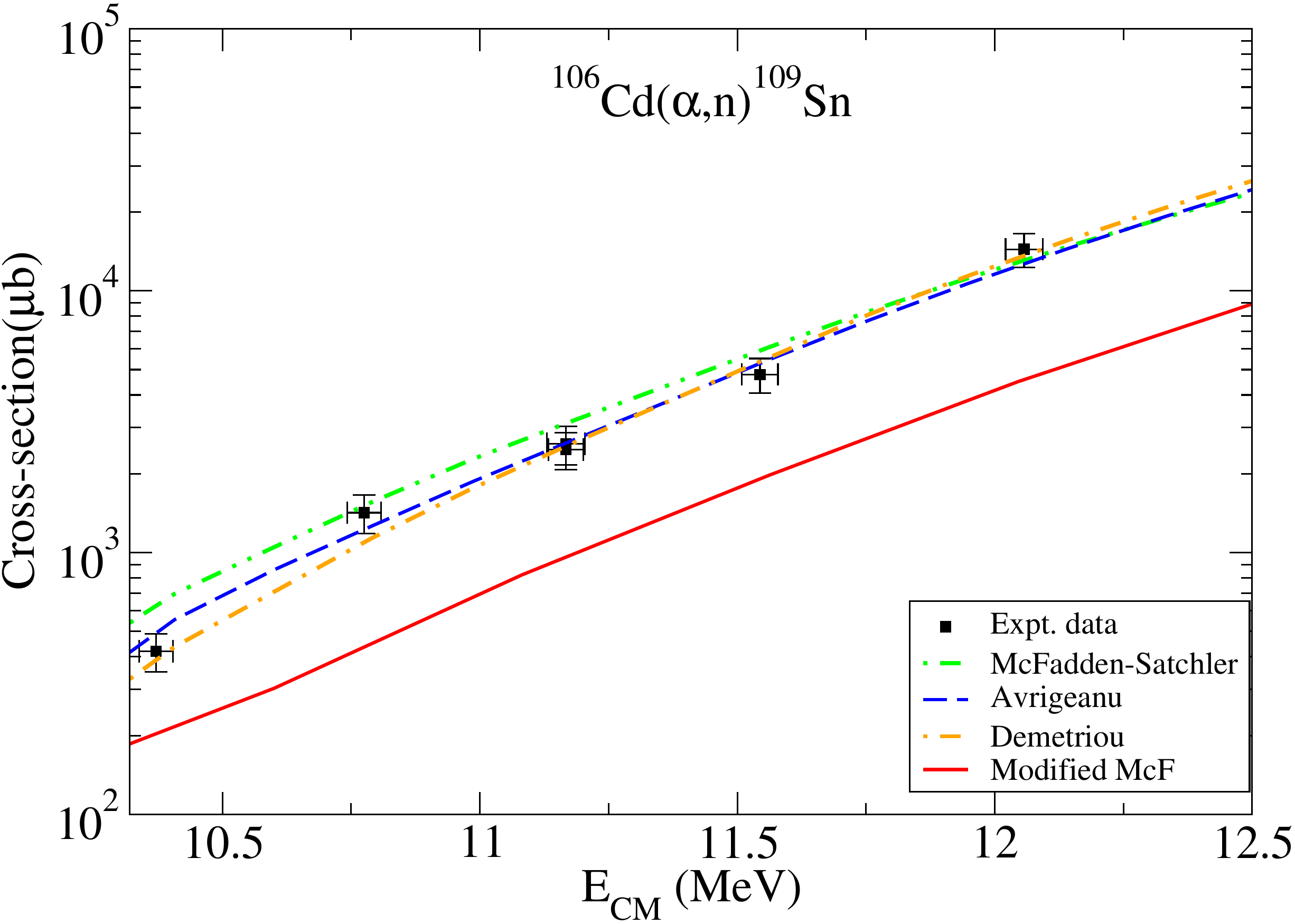}\\
\includegraphics[scale=0.24]{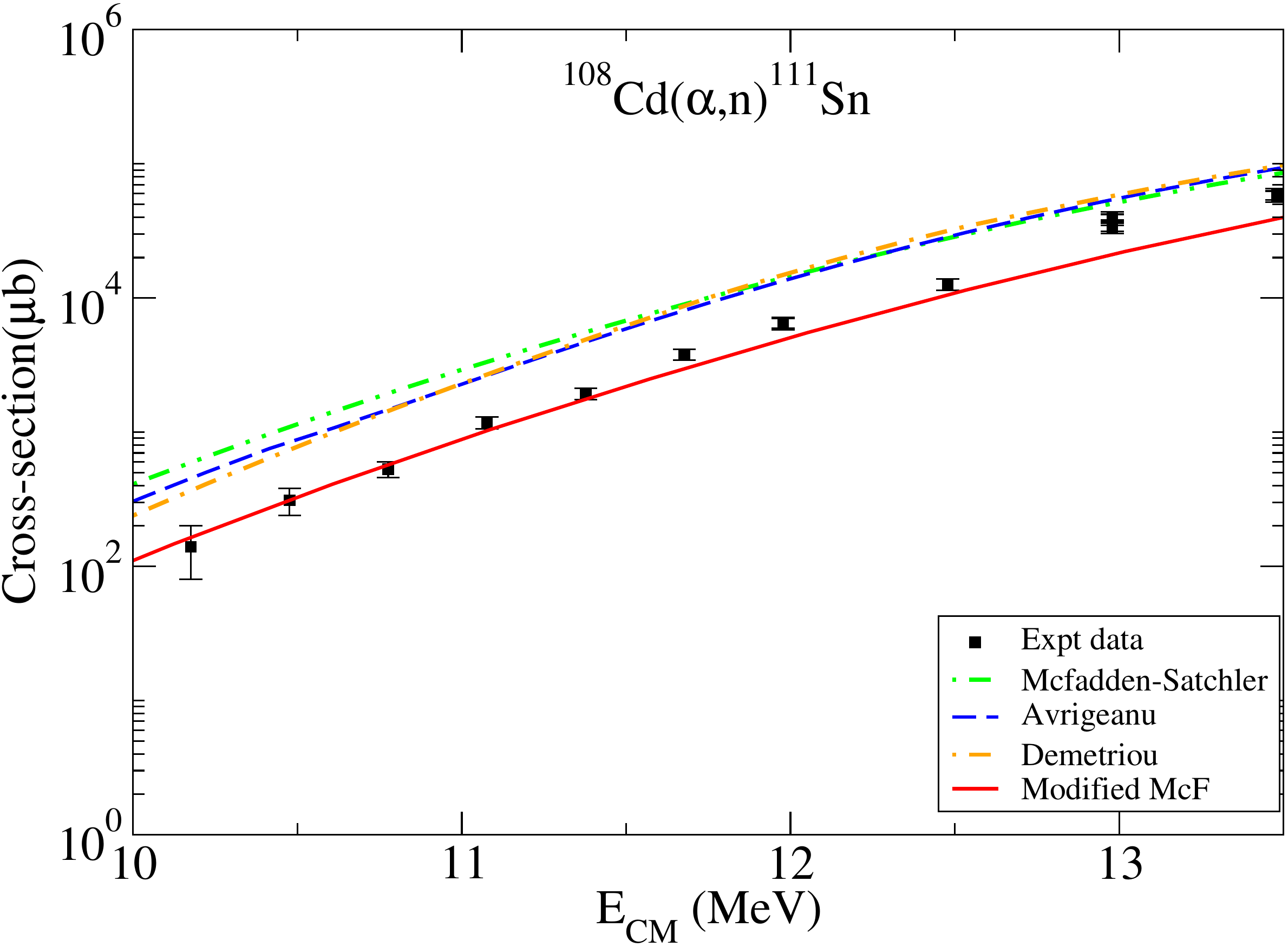}
\includegraphics[scale=0.24]{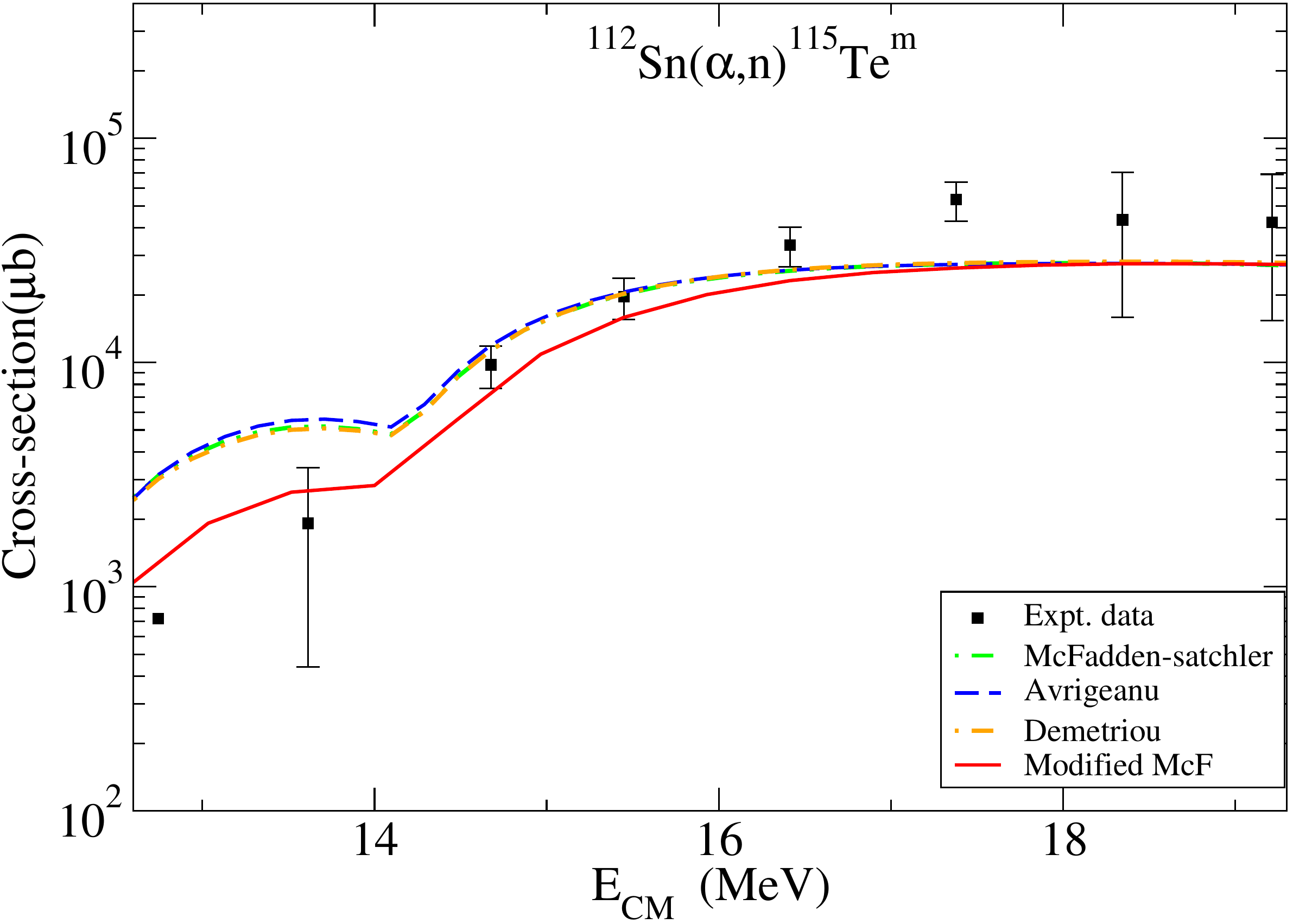}\\
\includegraphics[scale=0.24]{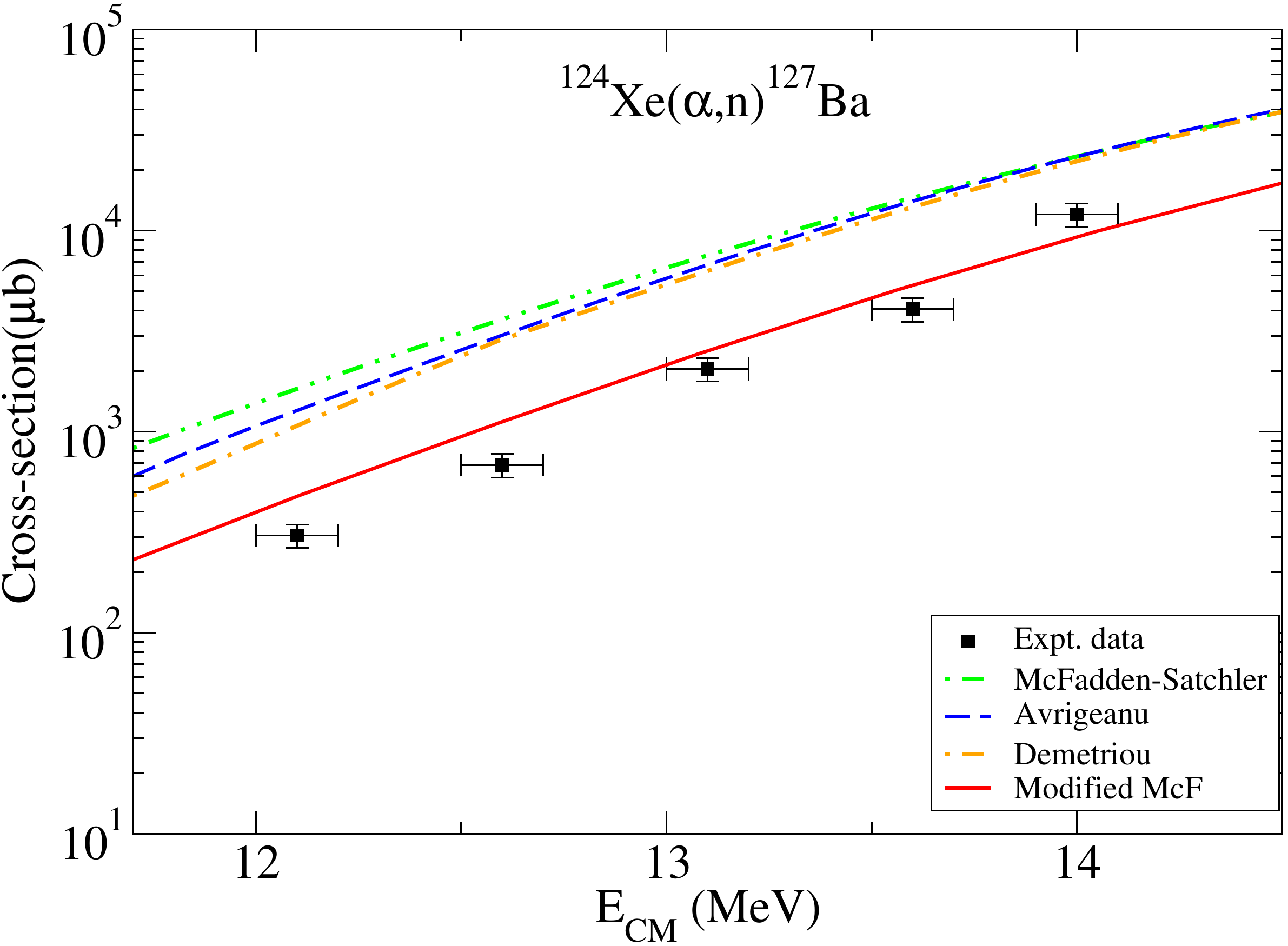}
\includegraphics[scale=0.24]{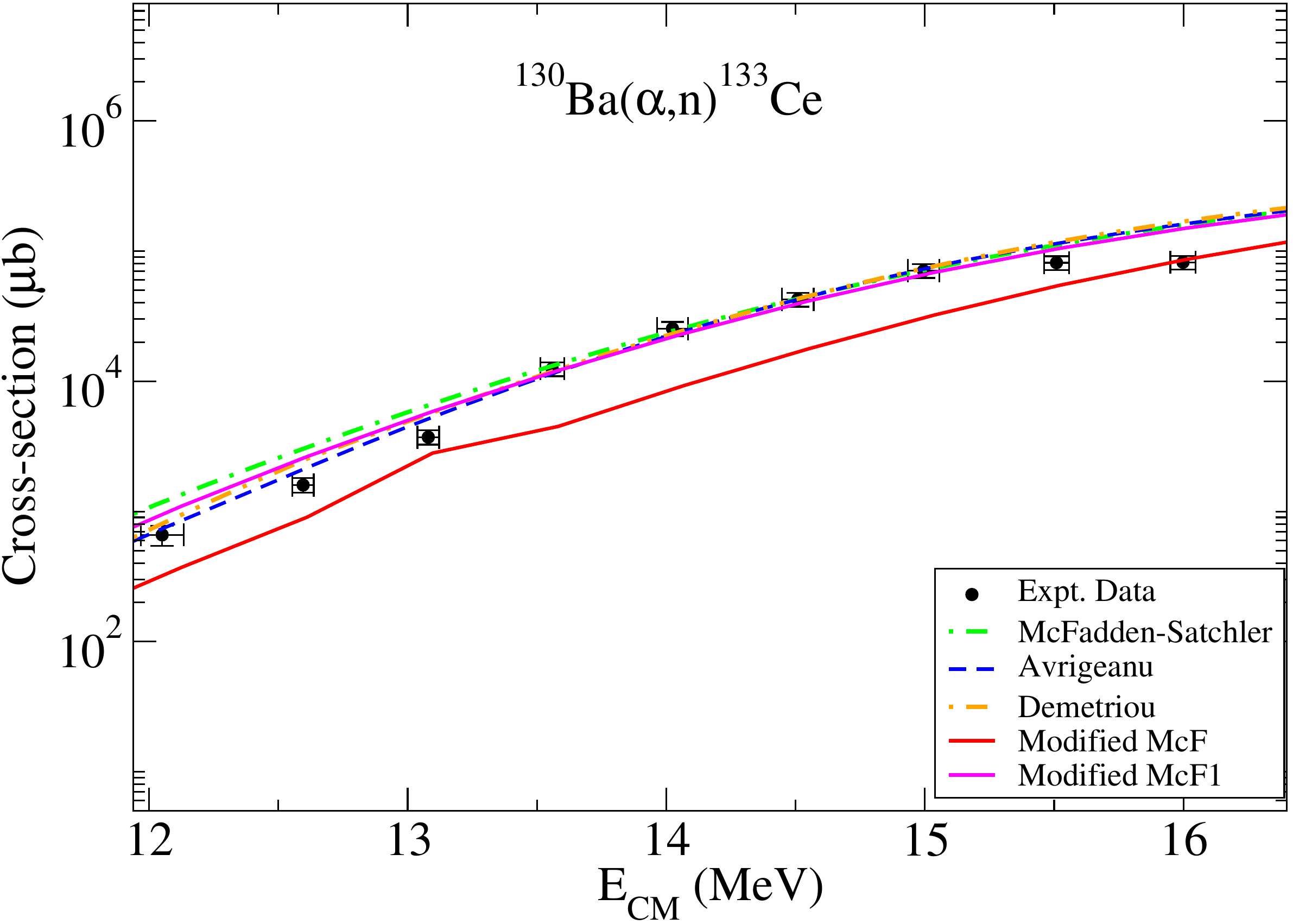}\\
\includegraphics[scale=0.24]{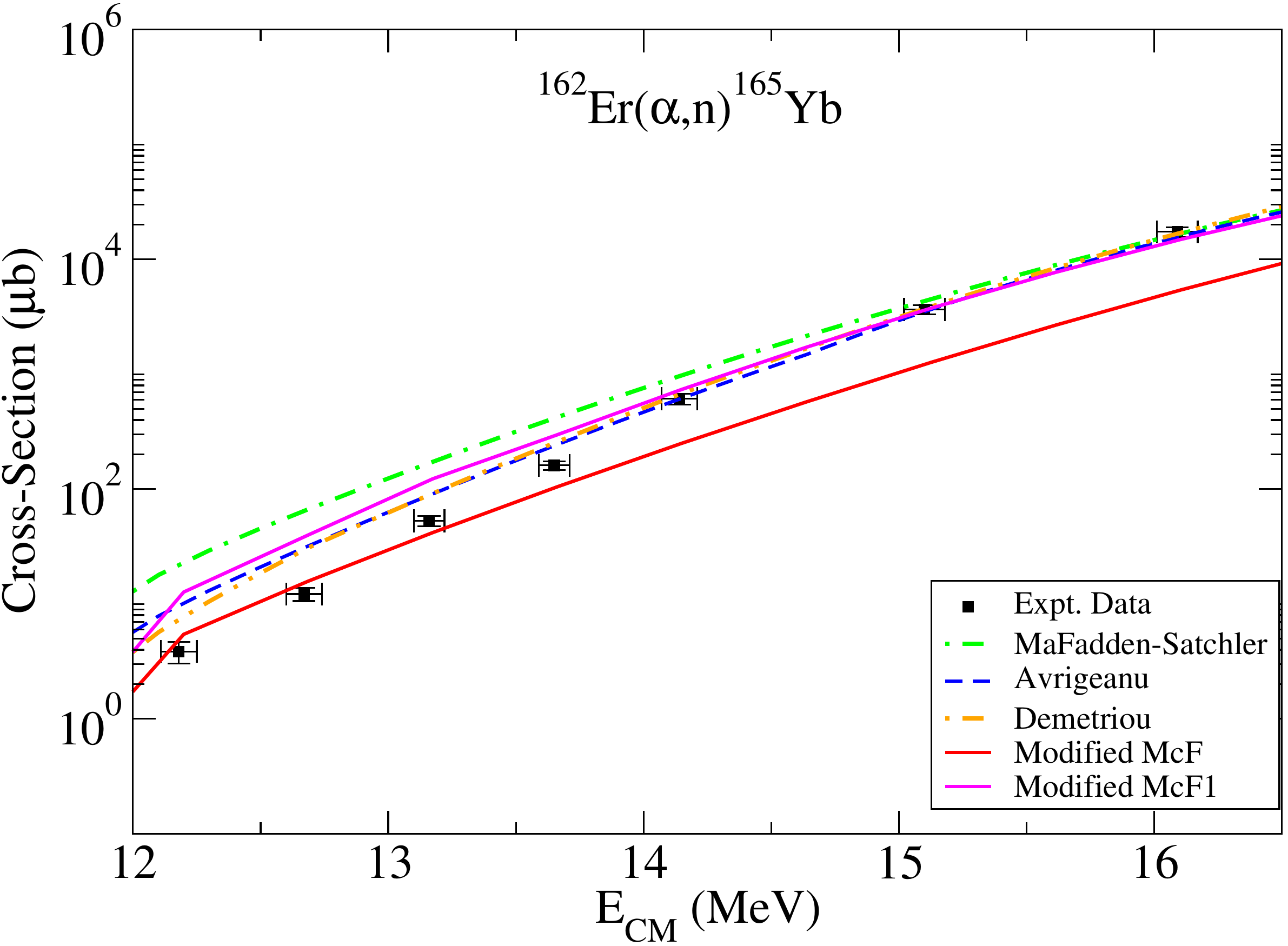}
\includegraphics[scale=0.24]{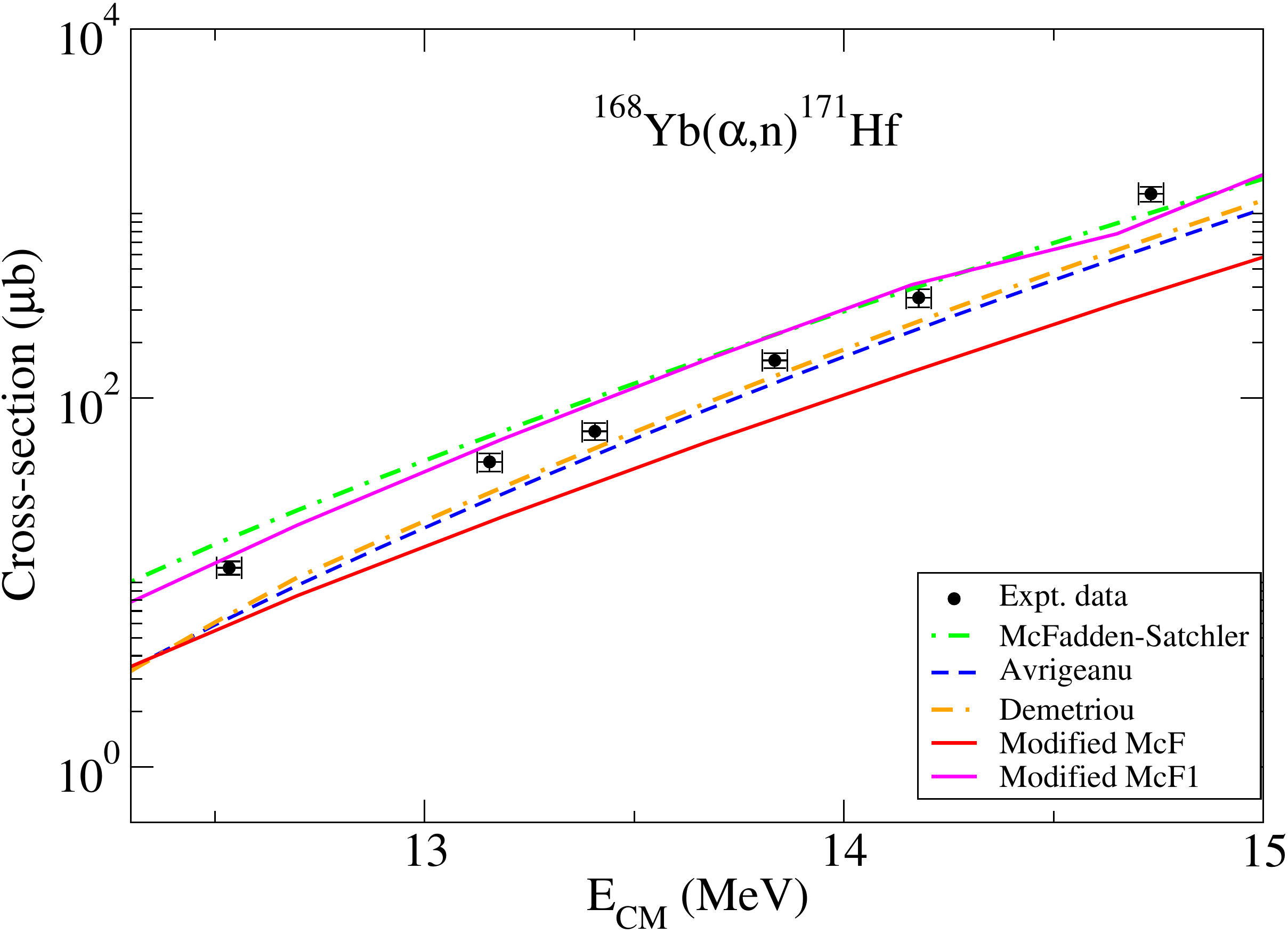}\\
\end{tabular}
\end{center}
\caption{($\alpha$,n) reaction cross-section of $^{92,94}$Mo, $^{96}$Ru, $^{106,108}$Cd, $^{112}$Sn, $^{124}$Xe, $^{130}$Ba, $^{162}$Er, and $^{168}$Yb. Experimental data from\cite{r,s,u,x,y,h1,i1,j1,k1} compared to HF calculation with different $\alpha$-optical potential.}
\label{FIG:2}       
\end{figure*}

\begin{table}[h!]
 \begin{center}
 \caption{Modified McF potential parameters Set } \label{tbl1}
 \begin{tabular}{ll}
 \hline\noalign{\smallskip}
 Real Potential parameter & Imaginary Potential parameter\\
 \noalign{\smallskip}\hline\noalign{\smallskip}
  $V_0 = \frac{185}{1+exp\left(\frac{0.9E_{cu} - E_{cm}}{a_c}\right)}$ & $W_v = \frac{25}{1+exp\left(\frac{0.9E_{cu} - E_{cm}}{a_c}\right)}$\\
  $r_v = 1.40$ &  $r_w = 1.40$ \\
  $a_v = 0.52$ & $a_w = 0.52$ \\
  \noalign{\smallskip}\hline\noalign{\smallskip}
 $ r_c = 1.30 $ & \\
 \noalign{\smallskip}\hline
 \end{tabular}
 \end{center}
 \end{table}
 
 \begin{table}[h!]
  \begin{center}
  \caption{$\chi^2$/N Values of ($\alpha$,n) reaction using different optical potential}\label{tbl2}
   \begin{tabular}{p{0.07\textwidth}p{0.07\textwidth}lll} 
    \noalign{\smallskip}\hline
     p-nuclei & Modified McF & McF & Avrigeanu & Demetriou\\
     \noalign{\smallskip}\hline\noalign{\smallskip}
      $^{92}$Mo&13.99&188.85&138.82&44.82\\
       $^{94}$Mo&30.12&92.03&22.84&7.09\\
        $^{96}$Ru&11.83&1413.44&481.49&256.46\\
         $^{106}$Cd&15.65&2.083&0.57&2.71\\
          $^{108}$Cd&8.46&137.69&91.66&42.37\\
           $^{112}$Sn&1.64&1.91&2.16&1.82\\
            $^{124}$Xe&10.39&581.28&378.47&296.18\\
            $^{130}$Ba&15.91&19.40&10.23&7.05\\
            $^{162}$Er&23.03&336.42&36.99&8.52\\
            $^{168}$Yb&31.95&90.39&2.44&18.18\\
            $^{168}$Yb$^*$&7.97&\\
            \noalign{\smallskip}\hline
             \end{tabular}
             \end{center}
             \footnotesize{$^*$ Only modify the imaginary depth of McF potential (Modified McF1)}
             \end{table}

This modified McF AOMP that explains the ($\alpha$,n) cross-sections are given in Table.\ref{tbl1} and the fitted cross-sections using modified McF is shown in Fig.\ref{FIG:2}. In Table.\ref{tbl1} $E_{cu}$ is the Coulomb barrier energy and $a_c$ corresponds to the energy diffusivity parameter. The value of $a_c$ is varied between 2-12 and a better fit was obtained at $a_c$ =10. Similar modification of the AOMP was done previously \cite{k1,l1,m1,n1} but for only modifying the imaginary depth parameter with a different $a_c$ value. The prescription of modifying the imaginary part of the McF potential gives satisfactory results only in the heavier mass region(A $>$ 130). $\chi^2$/N values for using different AOMP are listed in Table.\ref{tbl2}. \\

 \subsection{($\alpha,\gamma$) reaction}
 These AOMP parameters  so obtained are used to calculate the ($\alpha,\gamma$) cross-section and are shown in Fig.\ref{FIG:3} in comparison to the experimental data. Corresponding $\chi^2$/N are listed in Table.\ref{tbl3}. In the statistical model, ($\alpha,\gamma$) cross-section depends on the $\gamma$-ray strength function and level density parameters (Eq.\ref{Eq6} and Eq.\ref{Eq7}) apart from AOMP.           
 TALYS includes 6 level density models and 8 types of $\gamma$-ray strength functions. Calculations were performed with different combinations of level density model and $\gamma$-ray strength functions while using the modified McF AOMP. 
 
 \begin{table} [h!]
  \begin{center}
  \caption{$\chi^2$/N Values of ($\alpha,\gamma$) reaction using different optical potential}\label{tbl3}
   \begin{tabular}{p{0.07\textwidth}p{0.09\textwidth}lll} 
    \noalign{\smallskip}\hline
     p-nuclei & Modified McF + BSFGM + BAL& McF & Avrigeanu & Demetriou\\
     \noalign{\smallskip}\hline\noalign{\smallskip}
      $^{92}$Mo&5.304&71.28&27.16&35.93\\
       $^{94}$Mo&2.073&65.45&17.81&13.08\\
        $^{96}$Ru&9.59&106.29&31.79&37.78\\
         $^{106}$Cd&15.20&57.17&33.36&43.68\\
          $^{108}$Cd&1.94&25.71&2.93&4.43\\
          $^{108}$Cd$^{\#}$&2.49&3.40&3.92&4.15\\
           $^{112}$Sn&2.37&13.21&1.11&3.22\\
           $^{112}$Sn$^{\#}$&21.22&764.12&30.06&26.58\\
            $^{124}$Xe&72.73&517.74&23.89&11.15\\
            $^{130}$Ba&25.61&29.94&38.01&54.48\\
            $^{162}$Er&41.78&55.09&45.14&53.12\\
            $^{162}$Er$^*$&7.14&\\
            $^{168}$Yb&32.64&28.82&31.76&33.03\\
            $^{168}$Yb$^*$&14.14&\\
            \noalign{\smallskip}\hline
             \end{tabular}
             \end{center}
             \footnotesize{$^{\#}$ Data taken from another Experiment \\
 $^*$ Only modify the imaginary depth of McF potential (Modified McF1)}
             \end{table}
             
  \begin{figure*}
\begin{center}
\begin{tabular}{cc}
\includegraphics[scale=0.23]{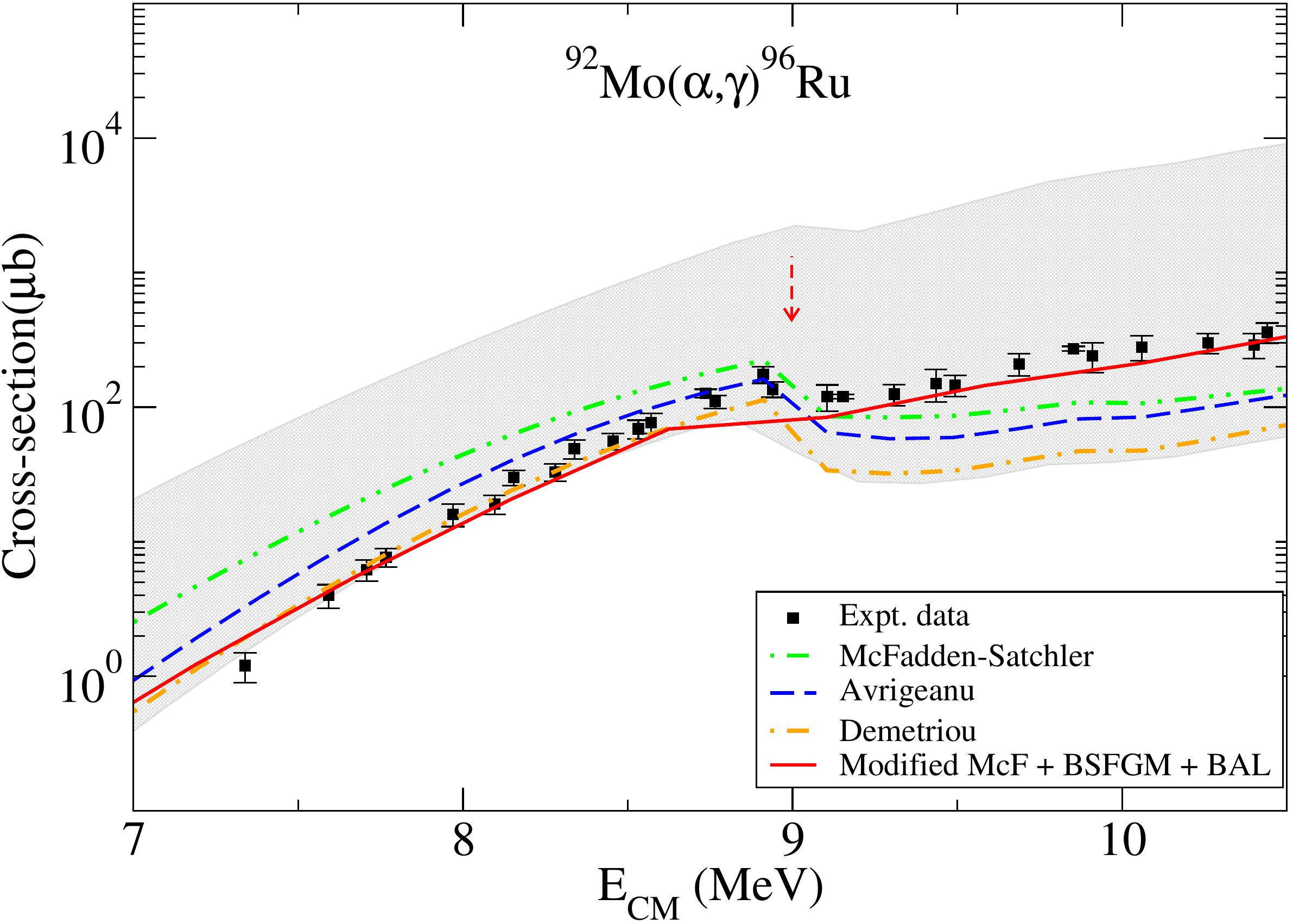}
\includegraphics[scale=0.23]{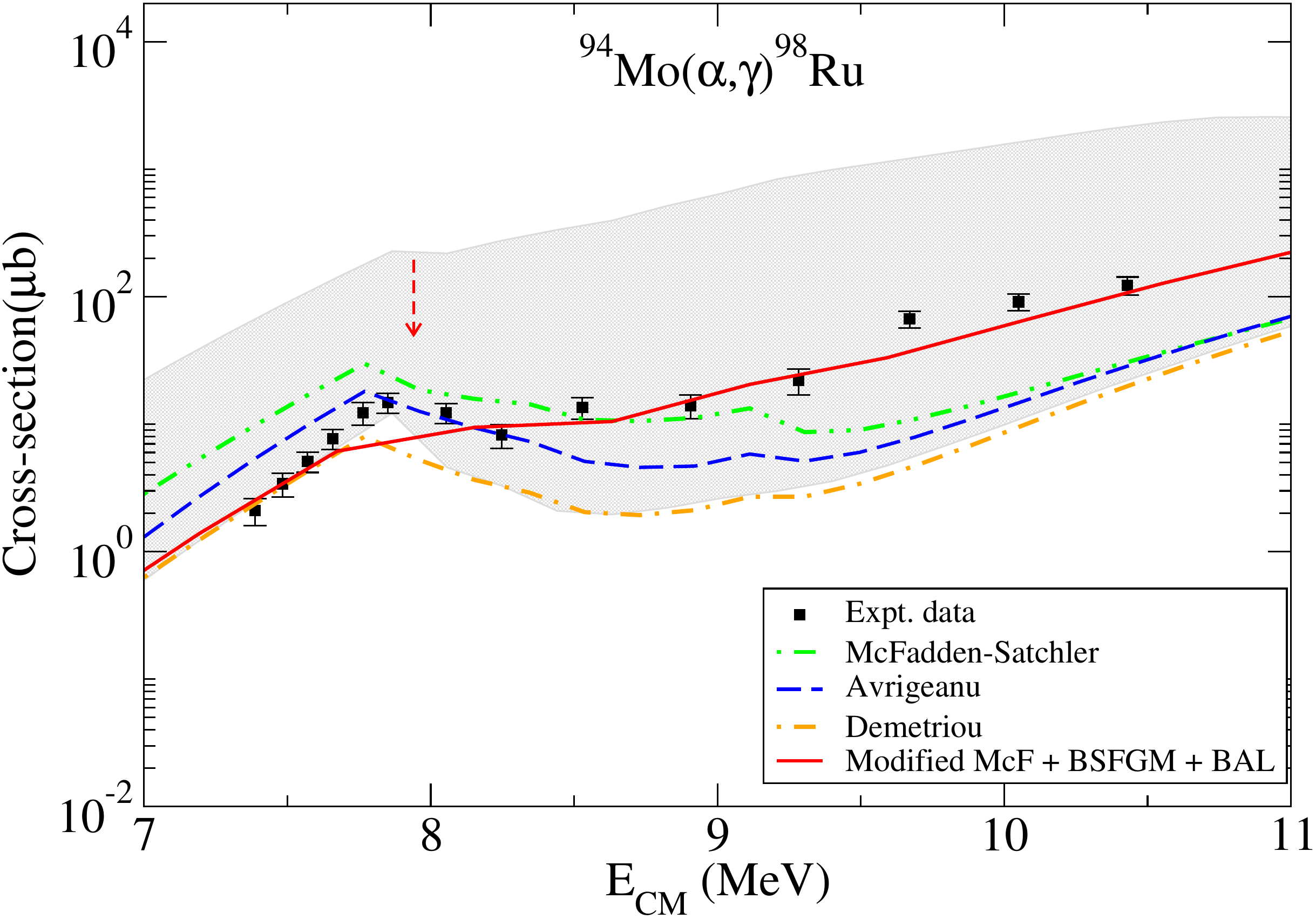}\\
\includegraphics[scale=0.23]{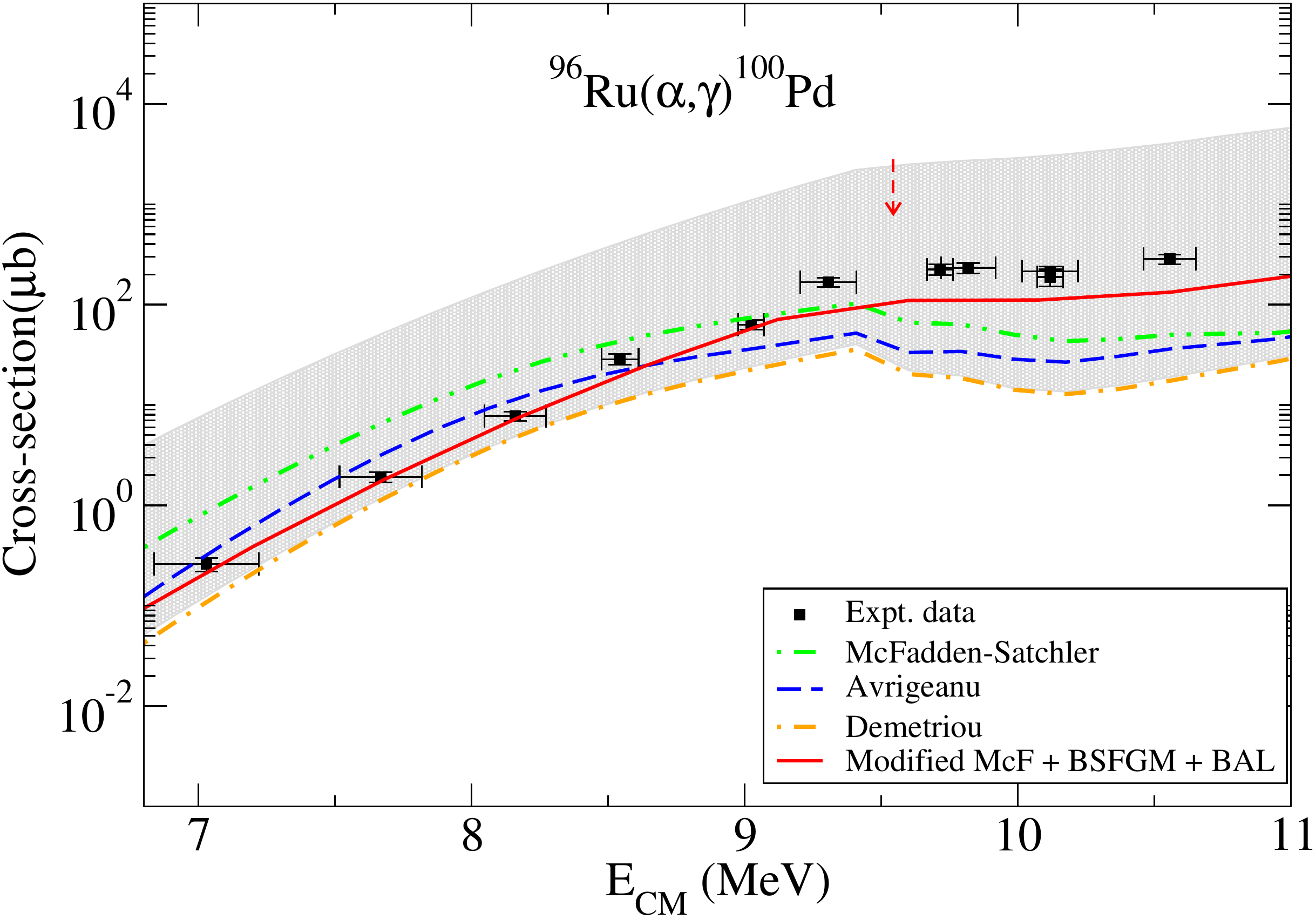}
\includegraphics[scale=0.23]{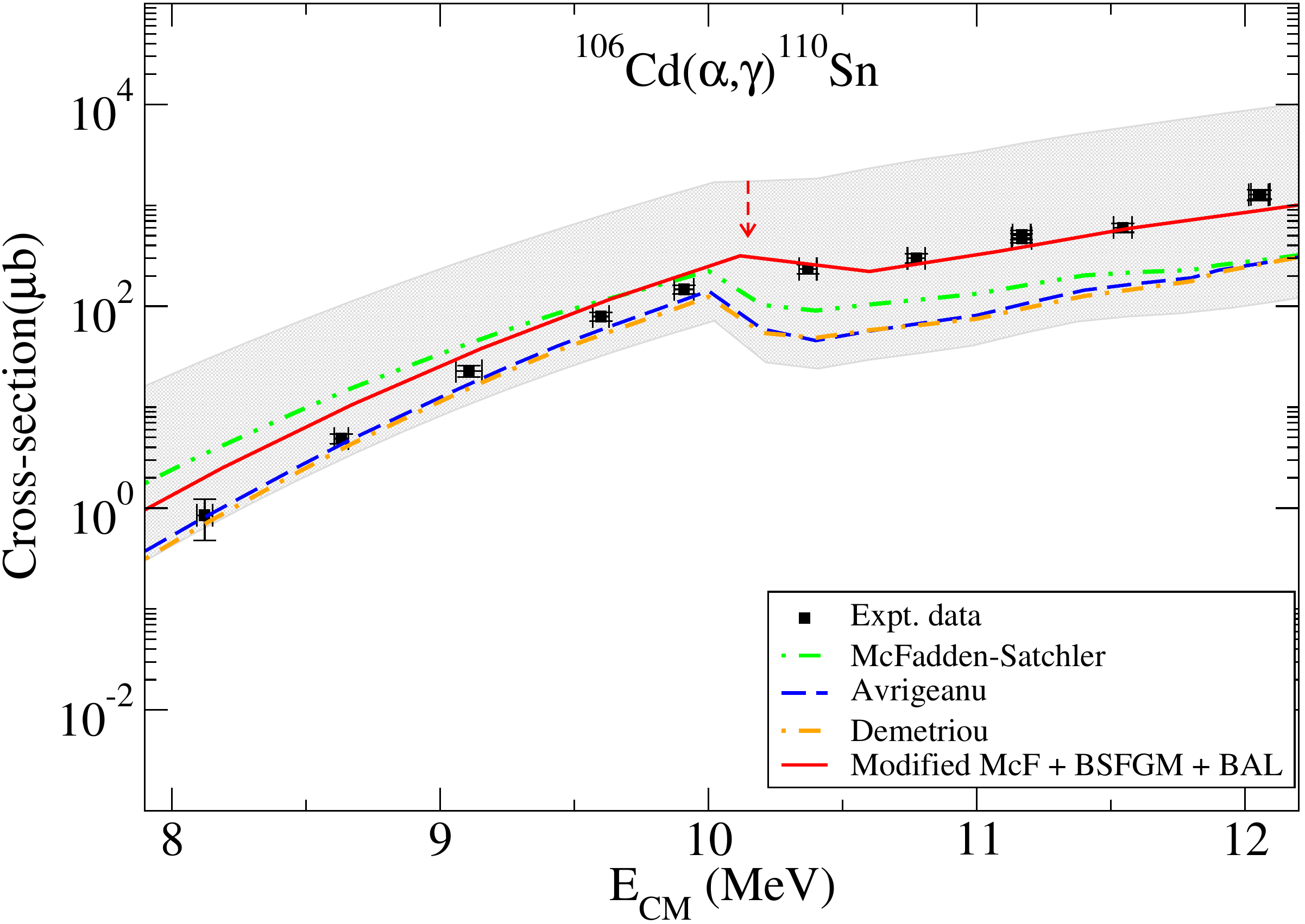}\\
\includegraphics[scale=0.23]{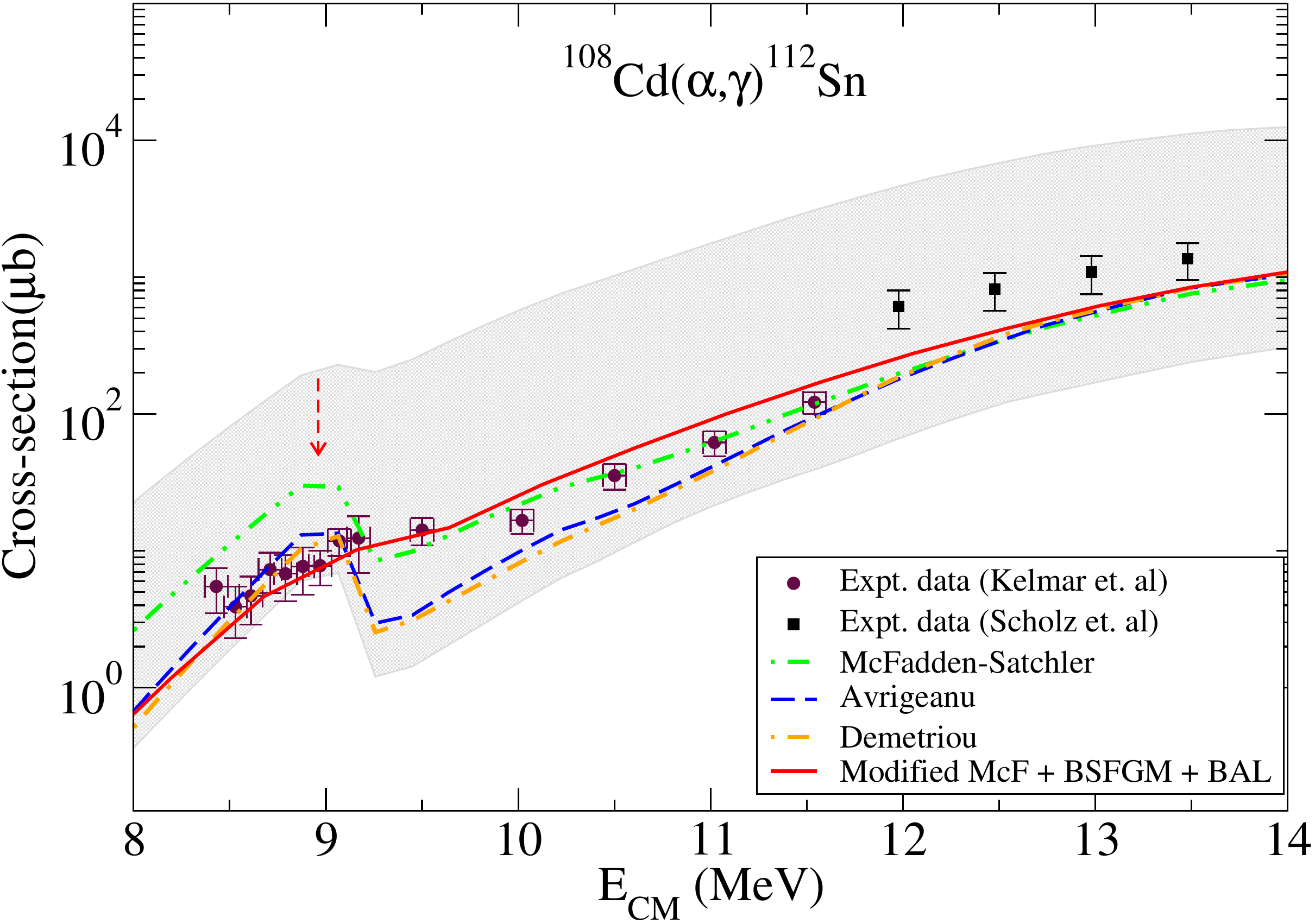}
\includegraphics[scale=0.23]{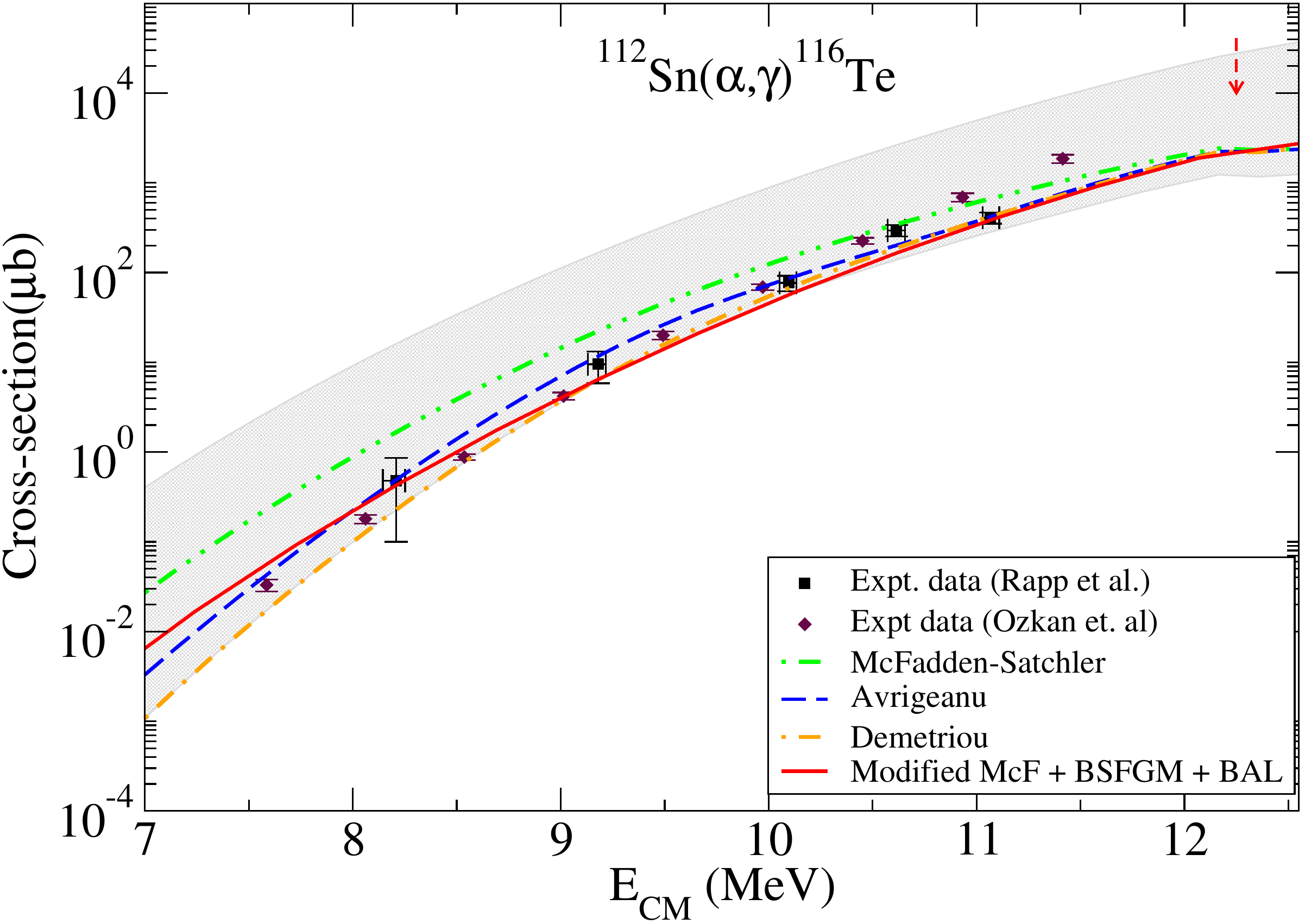}\\
\includegraphics[scale=0.23]{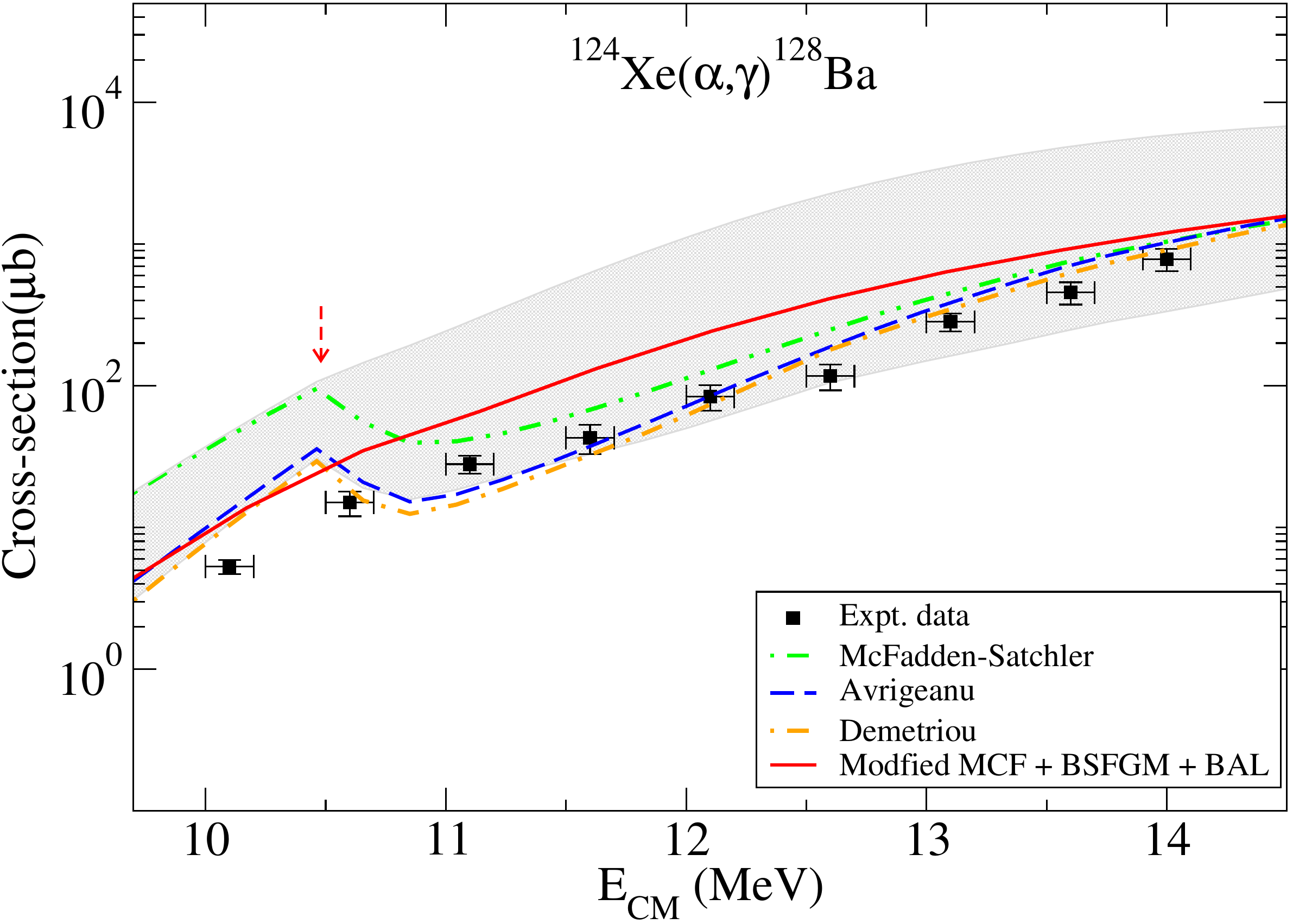}
\includegraphics[scale=0.23]{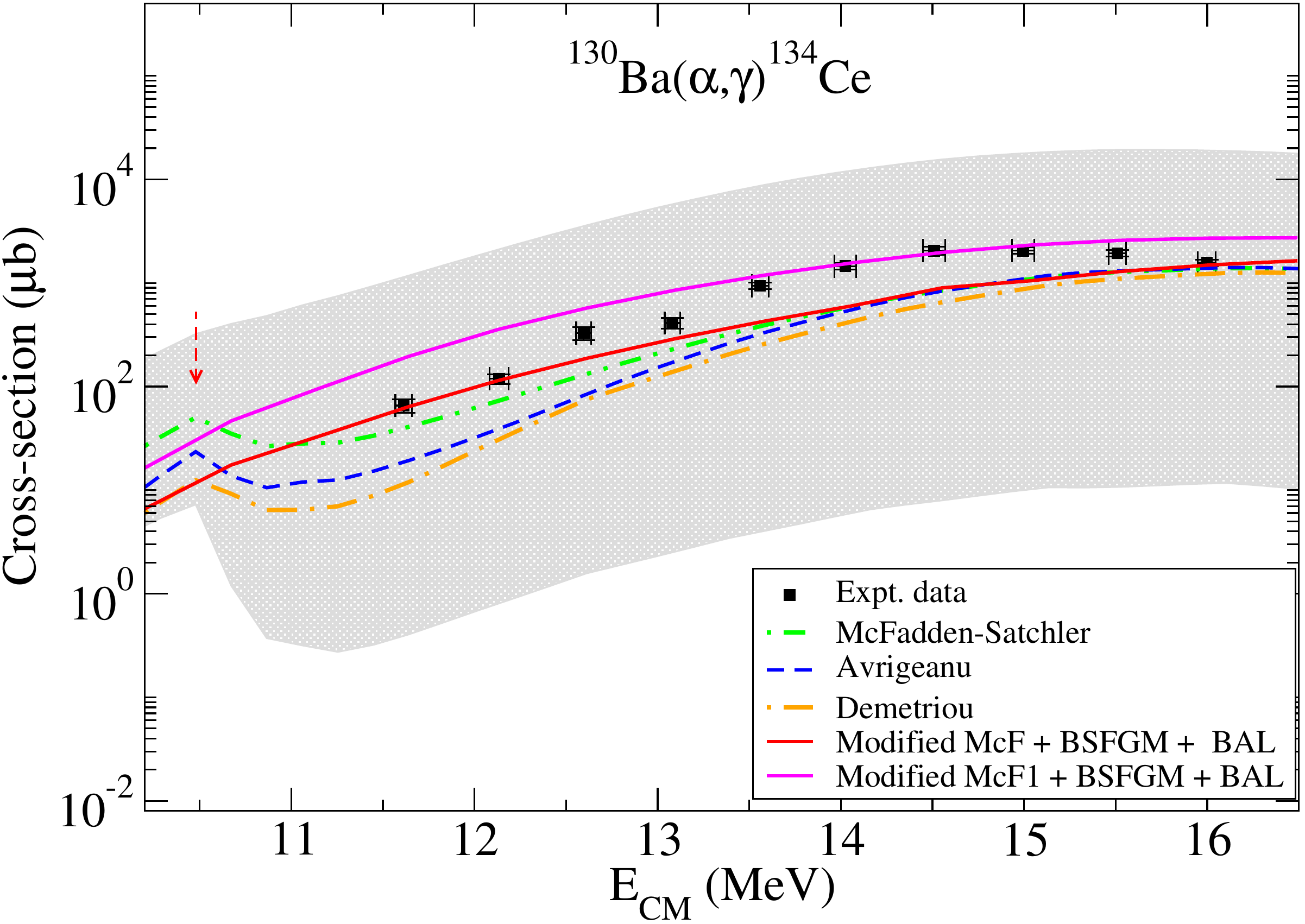}\\
\includegraphics[scale=0.23]{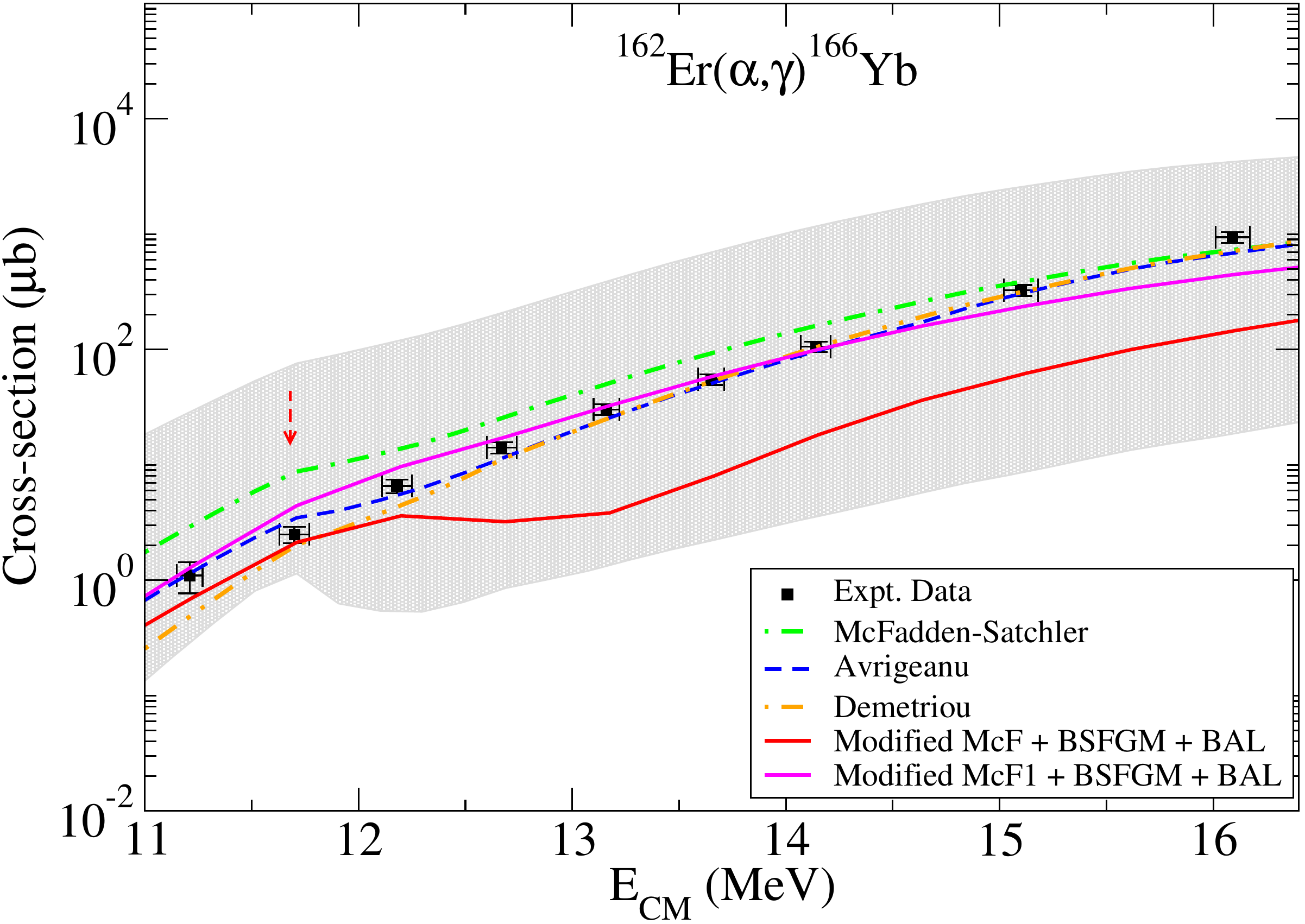}
\includegraphics[scale=0.23]{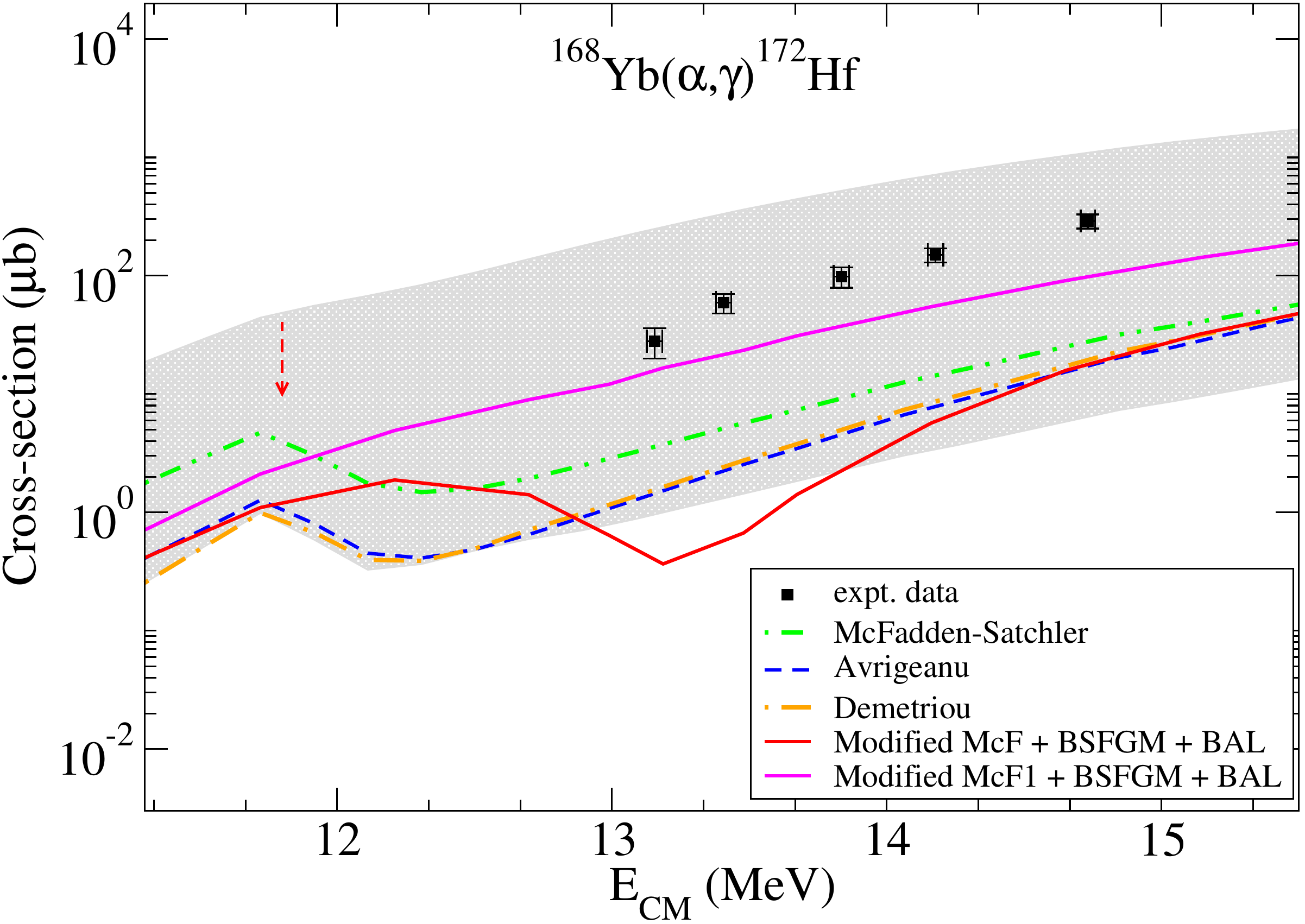}\\
\end{tabular}
\end{center}
\caption{($\alpha,\gamma$) reaction cross-section of $^{92,94}$Mo, $^{96}$Ru, $^{106,108}$Cd, $^{112}$Sn, $^{124}$Xe, $^{130}$Ba, $^{162}$Er, and $^{168}$Yb. Experimental data from\cite{r,s,u,v,w,x,y,i1,j1,k1,l1} compared to HF calculation with different $\alpha$-optical potential. Arrow implies the neutron threshold energy. The gray shaded region represents the predicted cross-section using different inputs (AOMP, LD, $\gamma$SF) in TALYS code.}
\label{FIG:3} 
\end{figure*}   

Alpha capture reaction cross-sections on p-nuclei are satisfactorily explained with the modified AOMP, Back-Shifted Fermi Gas model (BSFGM)\cite{o1} for level density and Brink-Axel Lorentzian(BAL)\cite{d1,e1} for $\gamma$-ray strength function.

\section{Summary and Conclusions}
  An alpha optical model potential (AOMP) is prescribed for use at astrophysical energies and particularly for reactions with p-nuclei. The potential is derived by using the ($\alpha$,n) reaction cross-section data at astrophysical energies, instead of using the conventional technique of ($\alpha,\alpha$) elastic scattering angular distribution data at much higher energies.\\
  The present AOMP is basically the well known McFadden and Satchler global alpha optical potential with its strength form modified. The general form of the new AOMP suggested for p-nuclei in terms of analysis of ($\alpha$,n) data is 
  \begin{equation}
  V_0 = \left(V_0\right)_{McF} \left[ 1 + F_1 c \right ]^{-1} ,\hspace{0.5cm}
  W_0 = \left(W_0\right)_{McF} \left[ 1 + F_2 c \right ]^{-1}
  \end{equation}
  Where $c$ = $exp\left(\frac{0.9E_{cu} - E_{cm}}{a_c}\right)$\\
  The values of $F_1$ and $F_2$ varies as a function of ($N-Z$) as follows 
  
  \begin{table}[h!]
  \begin{center}
  \caption{}
  \begin{tabular}{lll}
  \noalign{\smallskip}\hline
  (N-Z) &  $F_1$ & $F_2$\\
 \noalign{\smallskip}\hline\noalign{\smallskip}
  8-16($\neq$10)& 1&1\\
  18-28 & 0 & 1\\
  10 & 0 & 0\\
  \noalign{\smallskip}\hline
  \end{tabular}
   \end{center}
  \end{table}

This trend is observed on the basis of available data and can be further substantiated by further measurements over a broader mass range. The modified potential also provides a satisfactory representation of the ($\alpha,\gamma$) experimental data.


\begin{thebibliography}{*}
\bibitem{a} S.E. Woosley and W.M.Howard, ApJS {\bf 36}, (1978) 285.

\bibitem{b} M. Rayet et al., Astron. Asrophys. {\bf 227}, (1990) 271.

\bibitem{c} M. Arnould, S. Goriely, Physics Reports {\bf 384}, (2003) 1-84. 
\bibitem{d} T. Sauter and F. Käppeler, Physics Review C {\bf 55}, (1997) 3127.
\bibitem{e} J. Mayer et al., Physics Review C {\bf 93}, (2016) 045809.

\bibitem{f} V.  Foteinou et al.,  Eur. Phys. J. A {\bf 55}, (2019) (5) 67 .

\bibitem{g} Gy.Gyürky et al., Nuclear Physics A {\bf 922}, (2014) 112.

\bibitem{h} J. Bork et al., Physics Review C {\bf 58}, (1998) 524.

\bibitem{i} Bo Mei et al., Physics Review C {\bf 92}, (2015) 035803.

\bibitem{j} N. Ozkan et al. Nuclear Physics A {\bf 710}, (2002) 469.

\bibitem{k} I. Dillmann et al., Physics Review C {\bf 84}, (2011) 015802.

\bibitem{l} Gy Gyürky et al., Journal of Physics G {\bf 34}, (2007) 817.

\bibitem{m} Gy Gyürky et al., Journal of Physics G {\bf 34}, (2007) 817.

\bibitem{n} F.R. Chloupek et al., Nuclear Physics A {\bf 652}, (1999) 391.

\bibitem{o} Michael A. Famiano et al., Nuclear Physics A {\bf 802}, (2008) 26.

\bibitem{p} Michael R. T. Güray et al., Physics Review C {\bf 80}, (2009) 035804.

\bibitem{q} L. Netterdon et al., Physics Review C {\bf 90}, (2014) 035806.

\bibitem{r} W. Rapp et al., Physics Review C {\bf 66}, (2002) 015803.

\bibitem{s} Gy. Gyürky et al., Physics Review C {\bf 74}, (2006) 025805.

\bibitem{t} C. Yalçın et al., Physics Review C {\bf 79}, (2009) 0265801.

\bibitem{u} P.Scholz et al., Physics Letters B {\bf 761}, (2016) 247-252. 

\bibitem{v} R. Kelmar et al., Physics Review C {\bf 101}, (2020) 015801. 

\bibitem{w} N. Ozkan et al., Physics Review C {\bf 75}, (2007) 025801.

 \bibitem{x} W. Rapp et al., Physics Review C {\bf 78}, (2008) 025804.
 
 \bibitem{y} Z. Halasz et al., Physics Review C {\bf 94}, (2016) 045801.
 
\bibitem{z} T. Rauscher, Ap. J. Suppl. {\bf 201}, (2012) 26.

 \bibitem{a1} A.J. Koning, S. Hilaire, S. Goriely, \textit{TALYS 1.95 A nuclear reaction program}, 2019. 
 
 \bibitem{b1} V. Avrigeanu et al., Physics Review C {\bf 90},(2014) 044612.
 
  \bibitem{c1} A. Gilbert, A.G.W. Cameron, Canadian Journal of Physics {\bf 43}, (1965) 1446.
  
  \bibitem{d1} D.M. Brink, Nuclear Physics {\bf 4}, (1957) 215.

\bibitem{e1} P. Axel, Physical Review {\bf 126}, (1962) 671. 
 
 \bibitem{f1} L. McFadden, G.R. Satchler, Nuclear Physics {\bf 84}, (1966) 177.

\bibitem{g1} P. Demetriou et al., Nuclear Physics A {\bf 707}, (2002) 253.

 \bibitem{h1} D.E. Khulelidze et al., Soviet Physics, JETP {\bf 20}, (1965) 259.
 
  \bibitem{i1} P.Demetriou, et al., AIP Conference Proceedings {\bf 1090}, (2009) 293.
 
 \bibitem{j1} Z. Halasz et al., Physics Review C {\bf 85}, (2012) 025804.

\bibitem{k1} G.G.Kiss et al., Physics Letters B {\bf 735}, (2014) 40-44.

\bibitem{l1} L.Netterdon et al., Nuclear Physics A {\bf 916}, (2013) 149.

\bibitem{m1} E.Somorjai et al., Astron. Astrophys. {\bf 333}, (1998) 1112.

\bibitem{n1} A. Sauerwein et al., Physics Review C {\bf 84}, (2011) 045808. 


\bibitem{o1} W. Dilg et al., Nuclear Physics A {\bf 217}, (1973) 269.
 
\end{thebibliography}
\end{document}